\documentclass[man]{apa7}

\usepackage{lipsum}

\usepackage[american]{babel}
\usepackage{graphicx}
\usepackage{amssymb}
\usepackage{amsmath}
\usepackage{caption}
\usepackage{subfig}
\usepackage{hyperref}
\usepackage{xcolor}
\usepackage{booktabs}
\usepackage{url}
\usepackage{threeparttable}
\usepackage{multirow}
\usepackage{subcaption}
\usepackage{csquotes}
\usepackage{pifont}
\usepackage{float}
\newcommand{\cmark}{\ding{51}}%
\newcommand{\xmark}{\ding{55}}%

\usepackage[style=apa,sortcites=true,sorting=nyt,backend=biber]{biblatex}
\usepackage{tikz}
\usetikzlibrary{shapes.geometric, arrows, positioning}

\tikzstyle{process} = [rectangle, rounded corners, minimum width=2.8cm, minimum height=0.8cm, text centered, draw=black, fill=blue!10, font=\footnotesize]
\tikzstyle{arrow} = [thick,->,>=stealth]
\DeclareLanguageMapping{american}{american-apa}
\title{Complex Dynamics in Psychological Data: Mapping Individual Symptom Trajectories to Group-Level Patterns}
\shorttitle{}

\author{Eleonora Vitanza \textsuperscript{1}, Pietro De Lellis \textsuperscript{2}, Chiara Mocenni \textsuperscript{1}, Manuel Ruiz Marin \textsuperscript{3}}
\authorsaffiliations{1. University of Siena, 2. University of Naples Federico II, 3. Universidad Politécnica de Cartagena}

\leftheader{}
\abstract{This study integrates causal inference, graph analysis, temporal complexity measures, and machine learning to examine whether individual symptom trajectories can reveal meaningful diagnostic patterns.
Testing on a longitudinal dataset of $N=45$ individuals affected by General Anxiety Disorder (GAD) and/or Major Depressive Disorder (MDD) derived from \cite{fisher2017exploring}, we propose a novel pipeline for the analysis of the temporal dynamics of psychopathological symptoms. First, we employ the PCMCI+ algorithm with nonparametric independence test to determine the causal network of nonlinear dependencies between symptoms in individuals with different mental disorders. We found that the PCMCI+ effectively highlights the individual peculiarities of each symptom network, which could be leveraged towards personalized therapies. At the same time, aggregating the networks by diagnosis sheds light to disorder-specific causal mechanisms, in agreement with previous psychopathological literature. Then, we enrich the dataset by computing complexity-based measures (e.g. entropy, fractal dimension, recurrence) from the symptom time series, and feed it to a suitably selected machine learning algorithm to aid the diagnosis of each individual.
The new dataset yields $91\%$ accuracy in the classification of the symptom dynamics, proving to be an effective diagnostic support tool. Overall, these findings highlight how integrating causal modeling and temporal complexity can enhance diagnostic differentiation, offering a principled, data-driven foundation for both personalized assessment in clinical psychology and structural advances in psychological research.}


\authornote{
   \addORCIDlink{Corresponding author}{0000-0002-4229-3089}

  Correspondence concerning this article should be addressed to Eleonora Vitanza, Department of Information Engineering and Mathematics, University of Siena, Via Roma 56, Siena, 53100, Italy. E-mail: e.vitanza@phd.poliba.it}

\begin{document}
\maketitle
\section{Introduction}\label{sec:introduction}
\label{intro}


The conventional approach to the study of mental disorder is to seek for a common underlying cause of all symptoms, in analogy with medical work where the presence of a symptom is typically associated to the presence of a disease \parencite{hyland2011origins}. Such an approach to mental disorder is typically labelled latent variable, to recall the search for an invisible root cause of the disorder. Although latent variable models allowed to explain the effects of dynamic latent internalizing and externalizing factors in the development of comorbidity among common mental disorders \parencite{kessler2011effects}, they failed to capture the dynamic and interactive nature of symptom evolution \parencite{borsboom2011small}. As pointed out in recent literature, this is due to the absence of a central disease mechanism or pathogenic pathway in mental disorders \parencite{borsboom2013network, kendler2011kinds, kendler2012dappled}.

In response to these limitations, network analysis has gained traction as a promising alternative for studying the complexity of psychopathology \parencite{borsboom2017network, hofmann2018complex}. Unlike traditional approaches that treat symptoms as independent entities, network-based methods conceptualize mental disorders as systems of interacting symptoms that cause each other. Disentangling the causal mechanisms that underlie symptom activation allows for a more granular exploration of their dynamics \parencite{wichers2021narrative, bringmann2022psychopathological, contreras2019study}.

Network analysis provides several advantages over classical statistical models.
First, it enables researchers to identify key symptoms—so-called “bridge symptoms”—that link different disorder networks and contribute to comorbidity \parencite{groen2020comorbidity, ren2021network, cramer2010comorbidity}.
Second, it allows for the investigation of stable symptom structures, capturing how symptoms relate to each other within groups of individuals at a given point or over short time frames \parencite{burger2023reporting, borsboom2021network}.
Third, by leveraging intensive longitudinal data, network analysis can uncover temporal trajectories and dynamic interactions among symptoms, offering insight into how psychopathology evolves over time and supporting the development of adaptive, personalized interventions \parencite{piccirillo2022personalized, howe2020idiographic, epskamp2018personalized, bringmann2021person, fisher2019open}.

However, these approaches often rely on the assumption of linear relationships between symptoms, thereby neglecting the intricate nonlinear relationships between them \parencite{jeronimus2019dynamic, wright2020personalized}. Under the assumption of linearity, many approaches rely on correlation-based techniques, such as Structural Equation Modeling (SEM) \parencite{fisher2017exploring} or Vector Autoregression (VAR) \parencite{haslbeck2025testing}, which may only capture linear temporal dependencies \parencite{ullman2012structural, zivot2006vector}. This limitation is particularly problematic for understanding the complex feedback loops underlying psychopathologies, where symptom interactions may be strongly nonlinear \parencite{hosenfeld2015major}.

Moreover, network approaches typically focus on group-level analyses and overlook individual variability \parencite{haslbeck2022modeling, eaton2023review}. Additionally, current studies face challenges in generalizing individual-level findings to broader populations \parencite{mcnally2021network,fisher2018lack}. Heterogeneity in symptom expression raises concerns about whether individual symptom networks can meaningfully inform group-level patterns \parencite{hoekstra2023heterogeneity}. Addressing this challenge requires methodological innovations that can balance personalized symptom structures with generalizable insights applicable to clinical practice \parencite{borsboom2024integrating}.

Given these limitations in both latent variable and network approaches, our study offers an innovative approach that combines causal inference and machine learning techniques to provide a deeper understanding of symptom dynamics in psychopathology. While previous studies have explored idiographic symptom networks \parencite{fisher2017exploring} or group-level linear dependencies \parencite{haslbeck2025testing}, to the best of our knowledge no method has yet fully uncovered nonlinear dependencies in individual and group-level analysis in psychopathology.

This study addresses three main questions:
(1) How can we reconstruct individual-level, possibly nonlinear, causal interactions among symptoms?
(2) Do group-level patterns emerge despite individual heterogeneity, and can they help distinguish diagnostic groups?
(3) Can dynamic symptom features be used to classify diagnosis at the individual level?

We outline two complementary methodological pipelines (see Figure \ref{fig:pipeline}), designed to answer such distinct yet interrelated open questions. The first pipeline addresses the first two questions. Specifically, we elect the PCMCI+ algorithm to reconstruct causal interactions between symptoms. We first show its superiority in recovering nonlinear interactions on a toy model for which ground truth is available to then illustrate its effectiveness on real data. Then, we bridge individual-level analyses and group-level generalizations through aggregated fusion networks, which capture shared causal patterns across diagnostic groups. This step allows to distinguish different diagnostic groups.  

The second pipeline builds on the use of complexity-based measures extracted from temporal symptom dynamics. The use of these metrics boosts the performance of the supervised classification algorithm that we use to classify diagnosis at the individual level.

While these pipelines serve distinct analytical purposes, their outcomes are interdependent: causal patterns derived from idiographic networks can guide the interpretation of classification results, whereas diagnostic outcomes may, in turn, inform the refinement of causal models by highlighting symptom dynamics of clinical relevance.

\begin{figure}[h!]
    \centering
    \begin{tikzpicture}[node distance=1.2cm and 1.2cm]

    \node (data) [process] {Symptoms Time Series};
    \node (pcmci) [process] at (-4.5, -2) {Causal Discovery (PCMCI+ algorithm)};
    \node (fusion) [process, below=of pcmci] {Group-level (Fusion Networks)};
    \node (kernels) [process, below=of fusion] {Network Comparison (Graph Kernels)};
    \node (group) [process, below=of kernels] {Group-level Patterns};
    
    \node (features) [process] at (+4.5, -2) {Feature Extraction (Complexity measures)};
    \node (selection) [process, below=of features] {Feature Selection (Out Of Bag Importance)};
    \node (classifier) [process, below=of selection] {Diagnostic Classification (Bagged Trees Ensemble)};
    \node (clinical) [process, below=of classifier] {Clinical Decision Support};
    
    \draw [arrow] (data) -- (pcmci);
    \draw [arrow] (data) -- (features);
    
    \draw [arrow] (pcmci) -- (fusion);
    \draw [arrow] (fusion) -- (kernels);
    \draw [arrow] (kernels) -- (group);
    
    \draw [arrow] (features) -- (selection);
    \draw [arrow] (selection) -- (classifier);
    \draw [arrow] (classifier) -- (clinical);
    \draw [arrow, dashed] (group) -- (clinical);
    \draw [arrow, dashed] (clinical) -- (group);
    \end{tikzpicture}
    \vspace{0.5cm}
    \caption{Dual analysis pipeline for idiographic multivariate time series: the left branch performs causal discovery using PCMCI+ and group-level network aggregation, leading to pattern identification; the right branch extracts complexity measures, selects predictive features, and performs diagnostic classification for clinical decision support. Dashed bidirectional arrows indicate the complementary nature of the two analyses: causal insights support diagnostic interpretation, while diagnostic outcomes inform and refine causal understanding.}
    \label{fig:pipeline}
\end{figure}

To illustrate the viability of our analysis pipelines, we test them on a longitudinal dataset on idiographic symptom networks for Generalized Anxiety Disorder (GAD) and Major Depressive Disorder (MDD) \parencite{fisher2017exploring}. 
GAD and MDD are among the most prevalent and debilitating mental health conditions \parencite{izard2013patterns}, often co-occurring and sharing underlying mechanisms \parencite{dobson1985relationship, kalin2020critical, katon1991mixed}. 
While GAD is primarily characterized by excessive and uncontrollable worry, heightened physiological arousal, and cognitive hypervigilance, MDD manifests through persistent low mood, anhedonia, and dysregulation in reward-processing circuits \parencite{eysenck2006anxiety, kalin2020critical}.

The frequent comorbidity between these disorders suggests shared vulnerability factors; however, their distinct clinical presentations necessitate precise differentiation for accurate diagnosis and effective treatment \parencite{baldwin2002can, pollack2005comorbid}, making them an ideal test for our analysis pipelines. Indeed, the application to longitudinal data from subjects with GAD, MDD, or both disorders in comorbidity enables us to identify distinct patterns that differentiate the three groups. Specifically, our approach reveals the underlying causal relationships among symptoms while preserving individual heterogeneity, and addresses the challenging task of differential diagnosis among these closely related clinical presentations.

The remainder of the paper is structured as follows: Section \nameref{sec:methods} outlines the dataset and the methodological approaches used to address our research objectives; Section \nameref{sec:results} presents the primary findings of the work; and Section \nameref{sec:discussion} describes the implications, limitations, and future directions of the study.

\section{Methods}\label{sec:methods}
In this section, we describe the data and the methodological framework employed in our study. First, in Section \nameref{sec:data} we present the dataset provided by \textcite{fisher2017exploring}, which includes idiographic symptom time series from individuals diagnosed with GAD, MDD, or both.
Second, we outline the two complementary methodological pipelines of Figure \ref{fig:pipeline}. Specifically, in Section \nameref{sec:pcmci} we introduce the PCMCI+ algorithm, a recent and powerful method for causal discovery in multivariate time series. We have first validated its superiority over traditional methods—such as Vector Autoregression (VAR) and Transfer Entropy (TE)—using synthetic datasets with known linear and nonlinear causal structures. We also introduce the extension of PCMCI+ individual results to group-level analysis, enabling the comparison of symptom dynamics across diagnostic groups.
In Section \nameref{sec:graph_comparison} we describe the graph comparison techniques employed to analyze symptom dynamics and diagnostic group differences.
Lastly, in Section \nameref{sec:bagged_tree} we present the Bagged Tree algorithm used to classify individuals among diagnostic groups, including the feature extraction and selection procedures that support this analysis.

\subsection{Data}\label{sec:data}
The dataset used in this study was derived from the work of \textcite{fisher2017exploring}, which describes a longitudinal study in which participants with a primary diagnosis of mainly Generalized Anxiety Disorder (GAD), Major Depressive Disorder (MDD), or comorbid GAD and MDD completed intensive repeated measures assessments -- four per day for at least 30 days --  prior to therapy. This rich and detailed dataset offers unique opportunities for dynamic and person-specific analyses. In the original dataset (see \url{https://osf.io/8yadb/}), the primary diagnosis was the main focus, while other comorbidities were recorded but not central to the analysis. This approach was based on the assumption that the primary diagnosis would be the most relevant for studying the specific mechanisms underlying anxiety and depressive disorders \parencite{fisher2017exploring}.

For our study, we followed a similar approach but, at Fisher’s recommendation, expanded the sample by including five additional participants from a more comprehensive dataset provided directly by him, reaching 45 individuals. This expansion allowed for a more detailed exploration of symptom dynamics and increased the robustness of our findings.
To remain consistent with the original framework, we focused our analyses on the primary diagnosis only.
Furthermore, we focused specifically on distinguishing between GAD, MDD, and their comorbidity, as these represent the core diagnostic categories of interest. Therefore, participants with a primary diagnosis that included GAD, MDD, or both were assigned to one of these three categories, disregarding any additional comorbidity. For example, a participant diagnosed with GAD and agoraphobia was considered solely within the GAD group. 

In Table \ref{tab:participant_characteristics} there is a summary of participants’ characteristics and diagnostic information. Diagnoses were established through structured clinical interviews using the Anxiety and Related Disorders Interview Schedule for DSM-5, administered by trained graduate students under the supervision of a licensed clinical psychologist. The table also reports symptom severity scores based on the Hamilton Rating Scale for Anxiety (HAM-A) and the Hamilton Rating Scale for Depression (HAM-D). The HAM-A consists of 14 items rated from 0 to 4, yielding a total score range of 0–56, while the HAM-D includes 13 similarly rated items, with a total score range of 0–52. We refer to \textcite{fisher2017exploring} for more details about the assessment procedure.

\begin{table}[h]
    \centering
    \tiny
    \begin{threeparttable}
    \caption{Participants' characteristics: ID, sex, age, ethnicity, primary diagnosis, other comorbidities, Hamilton Rating Scale for Depression score (HAM-D), Hamilton Rating Scale for Anxiety score (HAM-A).} \label{tab:participant_characteristics}
    \begin{tabular}{llllllll}
        \toprule
        ID & Sex & Age & Ethnicity & Primary Diagnosis & Other Comorbidities & HAM-D & HAM-A \\
        \midrule
        P001 & Female & 28 & Latin & MDD,GAD & Panic & 23 & 27 \\
        P003 & Male & 29 & White & MDD,GAD & NA & 16 & 15 \\
        P004 & Female & 32 & Latina & GAD & NA & 16 & 33 \\
        P006 & Male & 26 & White & MDD,GAD & SAD & 13 & 13 \\
        P007 & Female & 33 & Black & MDD,GAD & Agor,SAD,SpecPhob & 11 & 17 \\
        P008 & Female & 23 & AsianAmerican & MDD,GAD & PTSD,BodyDys & 19 & 15 \\
        P009 & Female & 25 & Other & GAD,SAD & SpecPhob & 17 & 9 \\
        P010 & Male & 33 & AsianAmerican & MDD,GAD & SAD & 22 & 22 \\
        P012 & Female & 36 & Latin & GAD,Agor & NA & 9 & 13 \\
        P013 & Male & 26 & White & MDD,GAD & SAD & 14 & 19 \\
        P014 & Male & 22 & Latin & MDD & NA & 10 & 12 \\
        P019 & Female & 30 & AsianAmerican & MDD & SAD & 10 & 10 \\
        P021 & Male & 59 & Other & GAD & SAD & 15 & 16 \\
        P023 & Female & 64 & White & GAD & NA & 8 & 7 \\
        P025 & Male & 31 & White & GAD,SAD & NA & 15 & 14 \\
        P033 & Female & 28 & White & GAD & Agor,SAD,OCD & 8 & 14 \\
        P037 & Female & 28 & Latin & GAD,SAD & IllnessAnxiety,SpecPhob & 21 & 41 \\
        P040 & Female & 29 & White & GAD & Agor,SAD,MDD,SpecPhob & 21 & 41 \\
        P048 & Male & 57 & AsianAmerican & MDD,GAD & SAD,SpecPhob & 14 & 17 \\
        P068 & Female & 42 & White & GAD & NA & 11 & 14 \\
        P072 & Female & 38 & AsianAmerican & MDD & GAD & 15 & 13 \\
        P074 & Female & 56 & White & MDD & NA & 12 & 10 \\
        P075 & Female & 27 & AsianAmerican & GAD & NA & 18 & 23 \\
        P100 & Male & 31 & White & GAD & PTSD & 7 & 14 \\
        P111 & Female & 23 & AsianAmerican & GAD & Panic,SAD,PTSD & 18 & 15 \\
        P113 & Female & 46 & Black & GAD & SAD,SpecPhob & 4 & 15 \\
        P115 & Female & 42 & White & MDD,GAD & SAD & 18 & 19 \\
        P117 & Male & 59 & White & MDD,GAD & NA & 12 & 18 \\
        P127 & Male & 29 & Latin & GAD & SAD & 9 & 13 \\
        P137 & Male & 45 & AsianAmerican & MDD & NA & 16 & 15 \\
        P139 & Female & 62 & White & MDD & GAD & 14 & 12 \\
        P145 & Female & 47 & Other & GAD & SAD,PTSD & 21 & 30 \\
        P160 & Male & 50 & White & GAD & PDD & 13 & 11 \\
        P163 & Female & 58 & AsianAmerican & MDD & GAD,PDD & 16 & 16 \\
        P169 & Male & 29 & White & MDD & NA & 13 & 15 \\
        P202 & Female & 34 & White & GAD & PDD & 10 & 11 \\
        P203 & Female & 21 & AsianAmerican & MDD,GAD & SAD & 18 & 20 \\
        P204 & Female & 57 & White & GAD & NA & 12 & 16 \\
        P206 & Female & 39 & Other & GAD,SAD & NA & 11 & 16 \\
        P215 & Female & 31 & Black & GAD & NA & 17 & 23 \\
        P217 & Female & 31 & White & GAD & MDD & 17 & 14 \\
        P219 & Female & 23 & AsianAmerican & GAD & MDD & 21 & 27 \\
        P220 & Male & 64 & White & MDD & GAD & 14 & 13 \\
        P223 & Male & 56 & White & MDD & GAD & 21 & 12 \\
        P244 & Female & 21 & AsianAmerican & MDD & NA & 12 & 8 \\
        \bottomrule
    \end{tabular}
    \begin{tablenotes}
            \scriptsize	
            \item Note. GAD = generalized anxiety disorder; MDD = major depressive disorder; Panic = panic disorder; Spec Phob = specific phobia; Agor = agoraphobia; PTSD = posttraumatic stress disorder; Body = body dysmorphia; AUD = alcohol use disorder; PDD = persistent depressive disorder.
        \end{tablenotes}
    \end{threeparttable}
\end{table}

Each participant completed an experience sampling protocol in which they responded to surveys delivered via mobile phone four times per day over a period of approximately 30 days. At each time point, participants rated their current experience of 22 mood and anxiety-related symptoms using a 0–100 visual analog slider with the anchors 'not at all' (0) and 'as much as possible' (100).
The symptom list included both core DSM-5 symptoms of GAD and MDD (e.g., felt worried, felt down or depressed, loss of interest or pleasure) and a broader set of affective, cognitive, and behavioral dimensions (e.g., felt energetic, felt hopeless, felt angry, avoided activities, felt accepted or supported; see Table \ref{tab:symptoms} for the full list). As a result, each participant's data can be represented as a time series with 22 variables (columns) capturing symptom dynamics over time. We further normalized the data column-wise by subtracting the mean and dividing by the standard deviation. Each individual completed an average of 112.8 responses (SD = 12.87), with a maximum of 151 and a minimum of 90. For each individual, entries with missing symptom values were excluded from the analysis. 
Aggregating data across individuals yielded $5076 \times 22$ observations, corresponding to 5076 time points and 22 symptom variables.
For each observation, we also included the participant ID and their primary diagnosis (GAD, MDD, or comorbid GAD and MDD), replicated across time points, resulting in a final dataset with $5076 \times 24$ entries, where each row corresponds to one survey response.

Although the diagnostic groups were unbalanced (GAD = 23, MDD = 11, comorbid = 11, see Table \ref{tab:group_descriptives} for complete group-level statistics), analyses were designed to be robust to class size differences. No resampling or weighting techniques were applied unless otherwise specified.
Treating comorbid GAD and MDD as a distinct group acknowledges growing evidence that symptom co-occurrence reflects unique dynamic patterns rather than additive effects of single disorders.

\begin{table}[ht]
    \centering
    \caption{List of the 22 symptoms assessed in the experience sampling protocol. Each item refers to how the participant felt or behaved during the time interval between the previous beep and the current one.}
    \label{tab:symptoms}
    \begin{tabular}{ll}
        \toprule
        \textbf{Symptom} \\
        \midrule
        Felt energetic & Felt hopeless \\
        Felt enthusiastic & Felt down or depressed \\
        Felt content & Felt positive overall \\
        Felt irritable & Felt fatigued or low energy \\
        Felt restless & Experienced muscle tension \\
        Felt worried & Had difficulty concentrating \\
        Felt worthless or guilty & Felt accepted or supported \\
        Felt frightened or afraid & Felt threatened, judged, or intimidated \\
        Loss of interest or pleasure & Dwelled on the past \\
        Felt angry & Avoided activities \\
        Procrastinated & Avoided people \\
        \bottomrule
    \end{tabular}
\end{table}

\begin{table}[ht]
    \centering
    \begin{threeparttable}
    \caption{Descriptive statistics by diagnostic group: number of individuals (N), mean (M) and standard deviation (SD) of Hamilton Rating Scale for Depression score (HAM-D), Hamilton Rating Scale for Anxiety score (HAM-A), age and responses to the experience sampling protocol.}
    \label{tab:group_descriptives}
    \begin{tabular}{lccc}
        \toprule
        \textbf{Variable} & \textbf{GAD} & \textbf{MDD} & \textbf{Comorbid GAD and MDD} \\
        \midrule
        N                    & 23 & 11 & 11 \\
        HAM-D (M ± SD)       & 13.87 ± 5.09 & 13.90 ± 3.14 & 16.36 ± 3.98 \\ 
        HAM-A (M ± SD)       & 18.70 ± 9.55 & 12.36 ± 2.42 & 18.36 ± 3.82 \\
        Age (M ± SD)         & 36.61 ± 11.86 & 43.72 ± 16.37 & 34.27 ± 13.03 \\
        Responses (M ± SD)   & 112.13 ± 13.74 & 109.09 ± 8.22 & 117.90 ± 14.17 \\
        \bottomrule
    \end{tabular}
    \begin{tablenotes}
        \small
        \item Note. GAD = generalized anxiety disorder; MDD = major depressive disorder.
    \end{tablenotes}
    \end{threeparttable}
\end{table}

Based on the primary diagnosis reported in Table \ref{tab:participant_characteristics}, all 45 participants were included in the analysis, comprising 29 females and 16 males, with ages ranging from 21 to 64 years (mean age = 37.8). The sample was ethnically diverse, including participants identifying as White (n = 20), Asian American (n = 12), Latin/Latina (n = 6), Black (n = 3), and Other (n = 4).

\subsection{Causal Inference through PCMCI+}\label{sec:pcmci}
To uncover causal relationships in multivariate time series, we adopted the PCMCI+ algorithm, a state-of-the-art method designed to infer both lagged and contemporaneous causal links in high-dimensional, autocorrelated, and potentially nonlinear datasets \parencite{runge2020discovering}. PCMCI+ is an optimized extension of the original PCMCI framework \parencite{runge2019detecting}. PCMCI itself combines the PC algorithm \parencite{spirtes2000causation}--- a constraint-based method for causal structure learning--- with an independence testing strategy tailored to time series data. PCMCI+ enhances this framework by refining the selection of conditioning sets for independence testing, thereby reducing false positives through a more robust pruning strategy.

We start by illustrating the PCMCI algorithm for causal discovery given the time series $y_1(t),\ldots,y_N(t)$ of the stochastic processes $Y_1(t),\ldots,Y_N(t)$, for $t=1,\ldots,T$. The PCMCI seeks the lagged dependencies between a target variable $Y_i(t)$ and a source variable $Y_j(t-\tau)$, with $j$ possibly also equal to $i$, and $\tau\le\tau_{\max}$ being a positive integer representing the lag. A graphical representation of a lagged dependency is a directed edge from the source to the target variable, with a label indicating the lag $\tau$. We denote the source $j$ as the parent of the target $i$. For all $\tau=1,\ldots,\tau_{\max}$, the PCMCI is performed in two steps:
\begin{enumerate}
    \item[($PC_1$)] For each variable $Y_i(t)$, a superset of potential lagged causal parents $\mathcal P^\tau(Y_i(t))$ of $Y_i(t)$ is identified through an iterative procedure. At the first iteration, $\mathcal P^\tau(Y_i(t))$ is composed by all the variables that fail the (unconditional) statistical independence test with $Y_i(t)$ at a given significance level $\alpha$, that is, checking $I(Y_j(t-\tau),Y_i(t))< \alpha$, with $I$ being the considered test statistic, see \cite{runge2019detecting} for details. Then, at every successive iteration $p > 1$, the set $\mathcal P^\tau(Y_i(t))$ is pruned by conditioning the independence test on a set $\mathcal{S}$ composed of the top $p$ strongest previously retained parents from $ \mathcal{P}^\tau(Y_i(t)) \setminus \{Y_j(t-\tau)\} $, ranked according to their test statistic. At the end of the algorithm, a potential parent $Y_j$ is retained in $\mathcal{P}$ if it is dependent of $Y_i$ when conditioned on $\mathcal{P}^\tau(Y_i(t))\setminus Y_j(t-\tau)$.
    Although this phase is designed to be sensitive and avoid false negatives, the resulting skeleton may still contain indirect or spurious connections. As discussed in \cite{runge2019detecting}, this is often due to limited statistical power, which depends on factors such as the sample size, the chosen significance level, the cardinality of the conditioning set, and the effect size—i.e., the magnitude of the conditional dependence measure $I(Y_i(t);Y_j(t-\tau)|\mathcal{S})$.
    \item[(MCI)] In the second stage, PCMCI applies the so called Momentary Conditional Independence (MCI) test to each edge identified in the first phase. This test verifies whether $Y_i(t)$ remains dependent on $Y_j(t-\tau)$ when, in addition to $\mathcal{P}^\tau(Y_i(t)) \setminus \{Y_j(t-\tau)\} $, we also condition on the subset of the $p_j$ strongest parents of $Y_j(t-\tau)$, with $p_j$ being a free parameter of the algorithm. The MCI phase is specifically designed to eliminate false positives resulting from indirect or mediated pathways, while retaining true direct causal influences.
\end{enumerate}
In its original formulation, PCMCI applies the MCI test only to lagged links. However, spurious or indirect links might appear also due to the presence of contemporaneous dependencies between variables. In this vein, the PCMCI+ algorithm, after performing a lagged conditioning phase as in the standard PCMCI, also includes a contemporaneous conditioning phase \parencite{runge2020discovering}. Once a more accurate skeleton has been obtained, PCMCI+ performs an orientation phase. This step is applied exclusively to contemporaneous links, as lagged links are naturally oriented by temporal order. Orientation is based on standard rules from the PC algorithm (the collider, propagation, and common child rule, see \cite{runge2020discovering},
which allow the algorithm to infer the direction of contemporaneous connections where possible.

Note that the PCMCI+ algorithm relies on standard assumptions from causal discovery theory. These include \textit{faithfulness}, which assumes that all observed independencies correspond to the absence of edges in the true causal graph; the \textit{Causal Markov Condition}, which states that each variable is independent of its non-effects (non-descendants) given its direct causes (parents); \textit{causal sufficiency}, which assumes that all relevant variables are observed and there are no hidden confounders; and \textit{stationarity}, meaning that the underlying causal mechanisms remain invariant over time.

Finally, we emphasize that a critical component of PCMCI+ is the conditional independence test employed. While linear methods like partial correlation are computationally efficient, they are limited in capturing nonlinear dependencies. To address this, we employed CMIknn \parencite{runge2018conditional}, a non-parametric estimator of conditional mutual information (CMI) based on $k$-nearest neighbors. Given the random variables $X$, $Y$, a vector of random variables $Z$, and $p(x,y,z)$ their joint mass probability function, CMI is formally defined as
\[
CMI:=I(X; Y \mid Z) = \sum_{x,y,z} p(x,y,z) \log \left( \frac{p(x,y \mid z)}{p(x \mid z) p(y \mid z)} \right).
\]
CMI quantifies the dependency between two variables $X$ and $Y$, conditioned on  $Z$, by estimating the reduction in uncertainty about $Y$ when $X$ is known, after accounting for $Z$. CMIknn provides a data-driven, distribution-free estimate of CMI, without assuming linearity, Gaussianity, or any specific parametric form.
This makes CMIknn particularly suitable for psychological time series, which often display nonlinear, noisy, and heterogeneous dynamics. Moreover, it is well-calibrated in finite-sample regimes and robust to a wide range of dependency structures. These features make it a flexible and powerful tool for uncovering meaningful, time-lagged symptom interactions in high-dimensional and idiographic datasets.
For instance, PCMCI+ has already shown promising results in biomedical domains, particularly in neuroscience. For example, it has been used to infer causal relationships between brain regions using hyperscanning EEG time series \parencite{silfwerbrand2024directed} and fMRI data \parencite{arab2025whole}, demonstrating its ability to extract meaningful temporal structures from complex biological signals.

\subsection{Applying the PCMCI+ to our dataset}
Importantly, we applied PCMCI+ at the level of each individual participant $i$. Each analysis was based on a time series of shape $T_i \times 22$, where $T_i$ denotes the number of observations associated to individual $i$. This individual-level approach allows for the construction of personalized causal graphs that reflect the intra-individual variability of symptom dynamics.
The analysis was configured with a maximum lag of one time step, as our primary interest lays in short-term rather than long-term trajectories, and with a significance level of $\alpha = 0.01$. 
We set $k = 4$ for the CMIknn estimator, as recommended for small sample sizes in previous studies \parencite{runge2018conditional}. This choice ensures a good balance between sensitivity and robustness.
All analyses were implemented using the Tigramite Python package (version 5.2.7.0), which provides a comprehensive framework for time series causal discovery, including full support for CMIknn. To ensure transparency and reproducibility, all code and data will be made available upon publication.


While causal discovery via PCMCI+ was conducted at the individual level—resulting in one causal graph per participant—
we also constructed group-level fusion networks by aggregating individual adjacency matrices within each diagnostic category (GAD, MDD, or comorbid). Specifically, we computed the element-wise sum of the binary causal graphs inferred by PCMCI+, yielding weighted matrices where each entry reflected the frequency of a specific edge within a group.

This aggregating approach, while methodologically distinct, aligns with prior work in network neuroscience \parencite{hagmann2008mapping}, where individual-level connectivity patterns were combined to identify common structural features across subjects. In our case, it enables the extraction of robust structural patterns that characterize each diagnostic profile, while preserving the idiographic nature of the analysis.

\subsection{Graph comparison}\label{sec:graph_comparison} 
To quantify the role of symptoms within each causal network, we first employed classical network measures from graph theory that capture both local and global aspects of node importance \parencite{freeman2002centrality}.
These metrics --- whose details are available in Section S3 of the Supplementary Information--- provide a multi-faceted view of symptoms importance to capture both local connectivity and global flow properties.

Secondly, to compare the graphs—both at the individual level and between fusion networks—we also employed graph kernels, a family of methods designed to quantify similarity between networks \parencite{kriege2020survey}. A graph kernel $K(G, G')$ is a positive semi-definite function that computes the similarity between two graphs $G$ and $G'$, with the key property that there exists a map $\phi$ from $G$ to a high-dimensional feature space such that $K(G, G') = \langle \phi(G), \phi(G') \rangle$ \parencite{vishwanathan2010graph}.

We considered two graph kernels differing in expressive power and computational complexity. For sparse graphs—such as the individual-level PCMCI+ outputs—we employed the \textit{degree distribution kernel}, which compares graphs by evaluating similarities in their node degree histograms \parencite{yaverouglu2015proper, sugiyama2018graphkernels}. In sparse networks, where variability in node connectivity is more pronounced, this kernel provides a computationally efficient and informative summary of the overall graph structure. Conversely, in denser graphs—where most nodes tend to exhibit uniformly high degrees—the degree kernel becomes less effective in distinguishing topological differences, as it captures less structural variability.
Thus, for comparing group-level networks, which are substantially denser than individual ones, we employed the \textit{Weisfeiler-Lehman (WL) kernel} \parencite{shervashidze2011weisfeiler}.

The WL kernel is a state-of-the-art method that assigns an initial color to each node and iteratively refines node colors by combining each node’s current color with those of its neighbors. After a fixed number of iterations--- e.g. when the kernel matrix converges--- each graph is represented by a histogram of node color counts, and similarity is computed as the dot product between these histograms. The WL kernel is particularly well suited for detecting structural differences in dense graphs, where local subgraph configurations carry greater discriminative power. While more computationally intensive, it provides higher expressiveness when structural complexity increases.

Importantly, in our case all the causal networks included the same set of nodes, ensuring that kernel comparisons were not affected by topological mismatches.
We computed kernel matrices both at the individual level (comparing all pairwise causal graphs) and at the group level (comparing fusion networks), enabling a multi-scale analysis of symptom dynamics. This allowed us to formally assess individual-level heterogeneity within and across diagnoses, as well as structural differences in the fusion networks.

\subsection{Classification through Bagged Tree}\label{sec:bagged_tree}
While the network-based analysis aimed at providing valuable insights into the structural patterns of symptom interactions at both individual and group levels, a complementary objective of our study was to explore whether temporal symptom dynamics could also support diagnostic classification. To this end, we implemented a supervised machine learning approach designed to leverage time series features for improving the accuracy and interpretability of diagnostic predictions in clinical psychology.

In this phase of analysis, we moved from an idiographic perspective to a global one, combining data from all individuals into a unified dataset. 
Data from individuals with comorbid GAD and MDD diagnoses were excluded from the training set and reserved for out-of-sample evaluation to assess model generalizability.
The final column of the dataset contained the diagnosis label (GAD or MDD), which served as the target variable for classification. 

We employed a Bagged Tree classifier—implemented in MATLAB using the \texttt{TreeBagger} function—as our predictive model \parencite{sutton2005classification}. Bagging, or bootstrap aggregation, is an ensemble learning method that improves predictive performance by reducing variance through model averaging \parencite{breiman1996bagging,breiman2017classification}. 
The model works in three steps: (1) random sampling with replacement creates diverse training sets, (2) individual decision trees are trained independently on these samples, and (3) predictions are aggregated via majority voting.
In order to account for diagnosis imbalance in the training set, we introduced class weights inversely proportional to class frequencies.

Our implementation consisted of 300 decision trees, each trained on a bootstrapped subset of the training data. All available predictors were considered at each split.

Although predictions were computed for each observation, individual-level classification was performed by averaging the predicted class probabilities across all time windows per participant.
We first assessed generalization using a leave-one-out cross-validation strategy, training the model on all individuals except one, who was used for testing. This simulates a real-world diagnostic setting in which a previously unseen participant is classified based on learned patterns from others. The final classification was based on an optimal decision threshold, determined by minimizing the Euclidean distance from the ideal point (0,1) in the ROC (Receiver Operating Characteristic) space—thereby balancing sensitivity and specificity.

To further enrich the analysis, we computed a comprehensive set of complexity features for each individual, designed to quantify different aspects of signal regularity, unpredictability, fractal geometry, and recurrence. 
Although they have not been widely applied in clinical psychology, prior studies in physiology and complex systems have shown that entropy, fractal dimension, and recurrence-based metrics can capture meaningful dynamic patterns in biological and behavioral data \parencite{goldberger2002fractal, kantz2003nonlinear, marwan2007recurrence, boccaletti2006complex}. 
The full list of complexity measures is available in Table~\ref{tab:complexity_metrics}, while detailed descriptions can be found in Section S4 of the Supplementary Information.

Thus, we obtained a high-dimensional dataset of approximately 3000 features. Each individual is represented by multiple rows in the dataset, corresponding to different configurations of parameters used to compute the complexity measures. Each column corresponds to a specific feature. We then trained the Bagged Tree classifier described above on this new enriched dataset.

Nonetheless, feature selection was performed post hoc leveraging out-of-bag (OOB) variable importance scores, specifically the Permuted Predictor Delta Error \parencite{guyon2003introduction, tang2014feature}. Variable importance scores were averaged across all iterations of the leave-one-out cross-validation to obtain a stable ranking of features contributing to diagnostic classification.
Features were then ranked in descending order of importance. Notably, some variables received negative importance values, indicating a detrimental effect on classification performance. To refine the feature set, we iteratively retrained the model, each time retaining only variables with positive importance. This Boruta-type process was repeated until only positively contributing variables remained and model performance reached a stable plateau \parencite{kursa2010feature}.

\begin{table}[ht]
    \centering
    \caption{Complexity metrics extracted from each time series or pairwise combination.}
    \label{tab:complexity_metrics}
    \begin{tabular}{ll}
        \toprule
        \textbf{Complexity metrics} & \\
        \midrule
        Number of Zero Crossings &
        Permutation Entropy \\
        Approximate Entropy &
        Sample Entropy \\
        Hurst Exponent &
        Detrended Fluctuation Analysis (DFA)\\
        Correlation Dimension &
        Higuchi Fractal Dimension \\
        Petrosian Fractal Dimension &
        Recurrence Rate \\
        Determinism &
        Laminarity \\
        Trapping Time &
        Maximum diagonal line length ($L_{max}$) \\
        Maximum vertical line length ($V_{max}$) &
        Divergence \\
        Entropy of Diagonal Line Lengths &
        Average Diagonal Length \\
        Average Vertical Length &
        Cross Recurrence Rate \\
        Cross Determinism &
        Cross Laminarity \\
        Cross Trapping Time &
        Cross $L_{max}$ \\
        Cross $V_{max}$ &
        Cross Divergence \\
        Cross Entropy of Diagonal Line Lengths &
        Cross Average Diagonal Length \\
        Cross Average Vertical Length &\\
        \bottomrule
    \end{tabular}
\end{table}

\section{Results}\label{sec:results}
\subsection{PCMCI+ outperformes classical approaches to inference from psychological data.}
To determine the most appropriate method for causal inference in longitudinal psychological data, we generated synthetic datasets designed to compare the ability of the PCMCI+ to reconstruct both linear and nonlinear dependencies, and compared it with Vector Autoregression (VAR) \parencite{zivot2006vector}, commonly used in the psychopathological literature \parencite{van2015ecological, haslbeck2025testing}, and transfer entropy (TE), a popular inference method from information theory \parencite{bossomaier2016transfer}\footnote{Further details on VAR and TE are available in Section S1 of the Supplementary Information.}. Each dataset included three variables, $X_1$, $X_2$, and $X_3$, observed over $100$ time points, with known ground-truth causal structures. The choice of relatively small number of datapoints aims to replicate a common limitation in idiographic psychological data, typically described by short time series.
Specifically, $X_{1,t}$ and $X_{2,t}$ were sampled independently from a standard normal distribution, $X_{i,t} \sim \mathcal{N}(0,1)$. The variable $X_{3,t}$ was generated according to three different causal scenarios, capturing either linear or nonlinear dependencies:

\begin{itemize}
\item Scenario 1 (Linear): \quad $X_{3,t} = 3 X_{2,t-1} + \varepsilon_{3,t}, \quad \varepsilon_{3,t} \sim \mathcal{N}(0,1);$
\item Scenario 2 (Interaction): \quad $X_{3,t} = 3 X_{1,t-1} X_{2,t-1} + \varepsilon_{3,t}, \quad \varepsilon_{3,t} \sim \mathcal{N}(0,1);$
\item Scenario 3 (Quadratic): \quad $X_{3,t} = 3 X_{2,t-1}^2 + \varepsilon_{3,t}, \quad \varepsilon_{3,t} \sim \mathcal{N}(0,1).$
\end{itemize}


The results of this comparison highlight the superior performance of the PCMCI+ algorithm—particularly when combined with the CMIknn independence test—in identifying both linear and nonlinear causal relationships, see Table~\ref{tab:toy_results}. Indeed, VAR and TE were restricted to detecting only linear dependencies. Interestingly, PCMCI+ with partial correlation also failed to recover nonlinear effects, emphasizing the value of nonparametric conditional independence testing in complex scenarios. 

The superior performance compared with the VAR model can be explained by the underlying linearity assumption. The case of transfer entropy is instead different, whereby it is a nonlinear method that quantifies how much knowing the past of a source variable reduces uncertainty about the future of a target variable. However, its main limitation lies in its data hungriness. Indeed, the computation of transfer entropy requires the discretization of the time series and the computation of empirical joint distributions. The accuracy of such estimation is critically limited for short time series \parencite{faes2013compensated}, as it is often the case for idiographic psychological datasets.

\begin{table}[ht]
\centering
\caption{Detection of linear and nonlinear causal relationships in synthetic datasets with three variables, evaluated across four methods: Vector Autoregression (VAR), Transfer Entropy (TE), PCMCI+ with partial correlation (parcorr), and PCMCI+ with the CMIknn independence test.}
\label{tab:toy_results}
\begin{tabular}{lcccc}
\toprule
\textbf{Causal relation} & \textbf{VAR} & \textbf{TE} & \textbf{PCMCI+ (parcorr)} & \textbf{PCMCI+ (CMIknn)} \\
\midrule
  $X_2 \rightarrow X_3$ & \cmark & \cmark & \cmark & \cmark \\
  $X_1 \cdot X_2 \rightarrow X_3$ & \xmark & \xmark & \xmark & \cmark \\
  $X_2^2 \rightarrow X_3$ & \xmark & \xmark & \xmark & \cmark \\
\bottomrule
\end{tabular}
\end{table}




\subsection{Group-level patterns in GAD, MDD, and comorbid participants emerge despite individual heterogeneity.}

\begin{figure}[H]
    \centering
    \includegraphics[width=\linewidth]{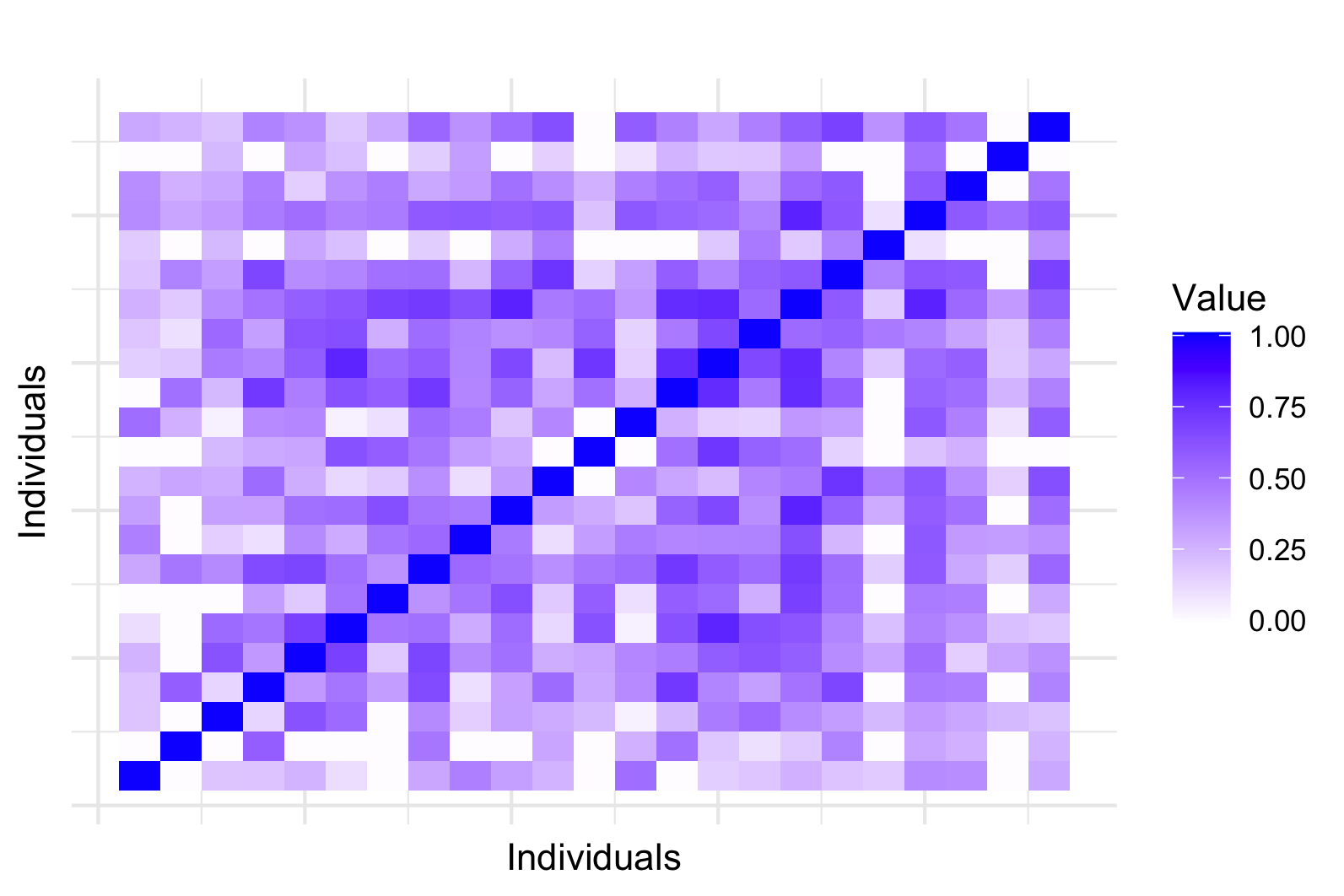}
    \caption{Degree kernel across individuals with Generalized Anxiety Disorder (GAD). The matrix shows the pairwise kernel similarity value among individual networks.}
    \label{fig:kernel_individuals_GAD}
\end{figure}

\begin{figure}[H]
    \centering
    \includegraphics[width=\linewidth]{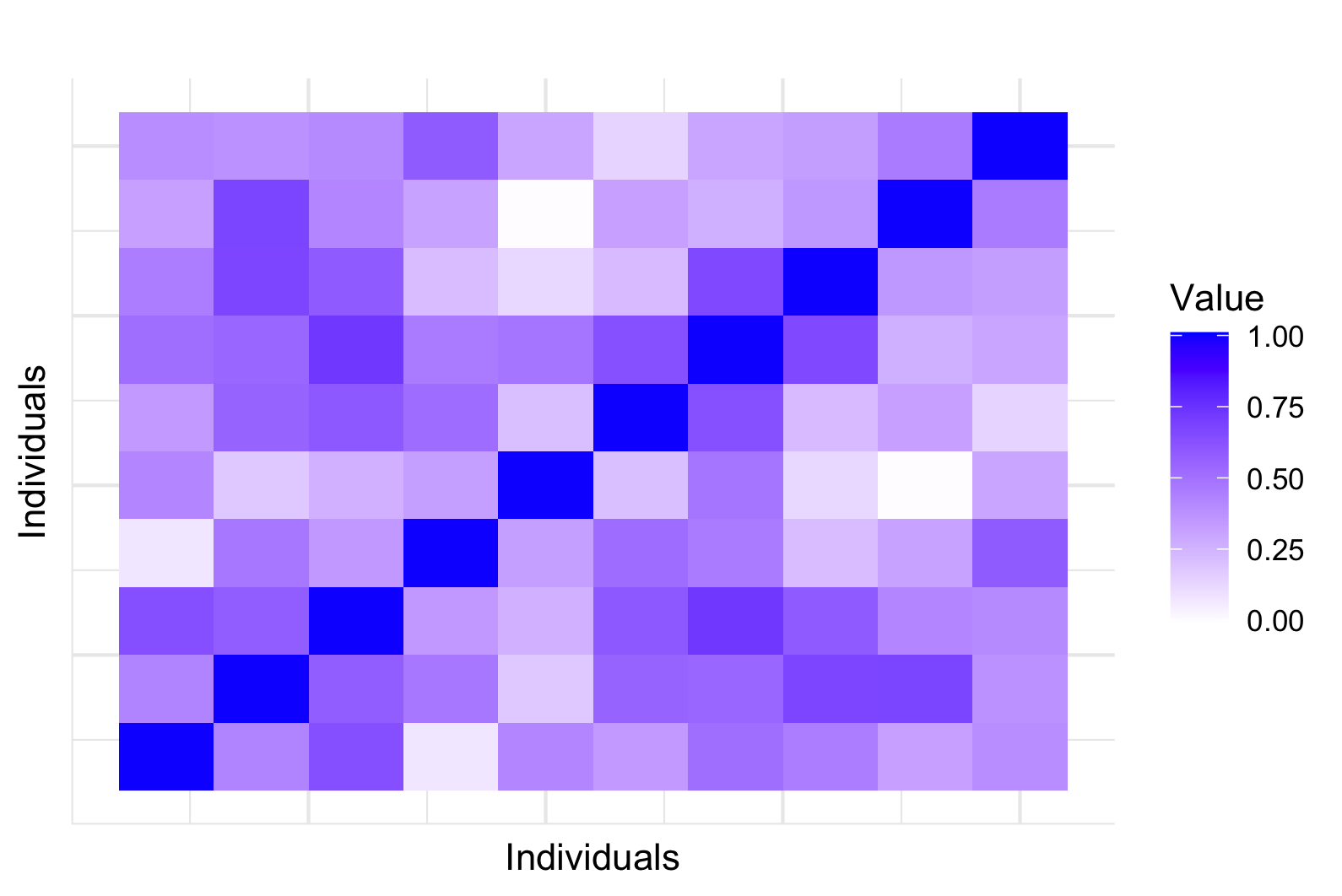}
    \caption{Degree kernel across individuals with Major Depressive Disorder (MDD). The matrix shows the pairwise kernel similarity value among individual networks.}
    \label{fig:kernel_individuals_MDD}
\end{figure}

\begin{figure}[H]
    \centering
    \includegraphics[width=\linewidth]{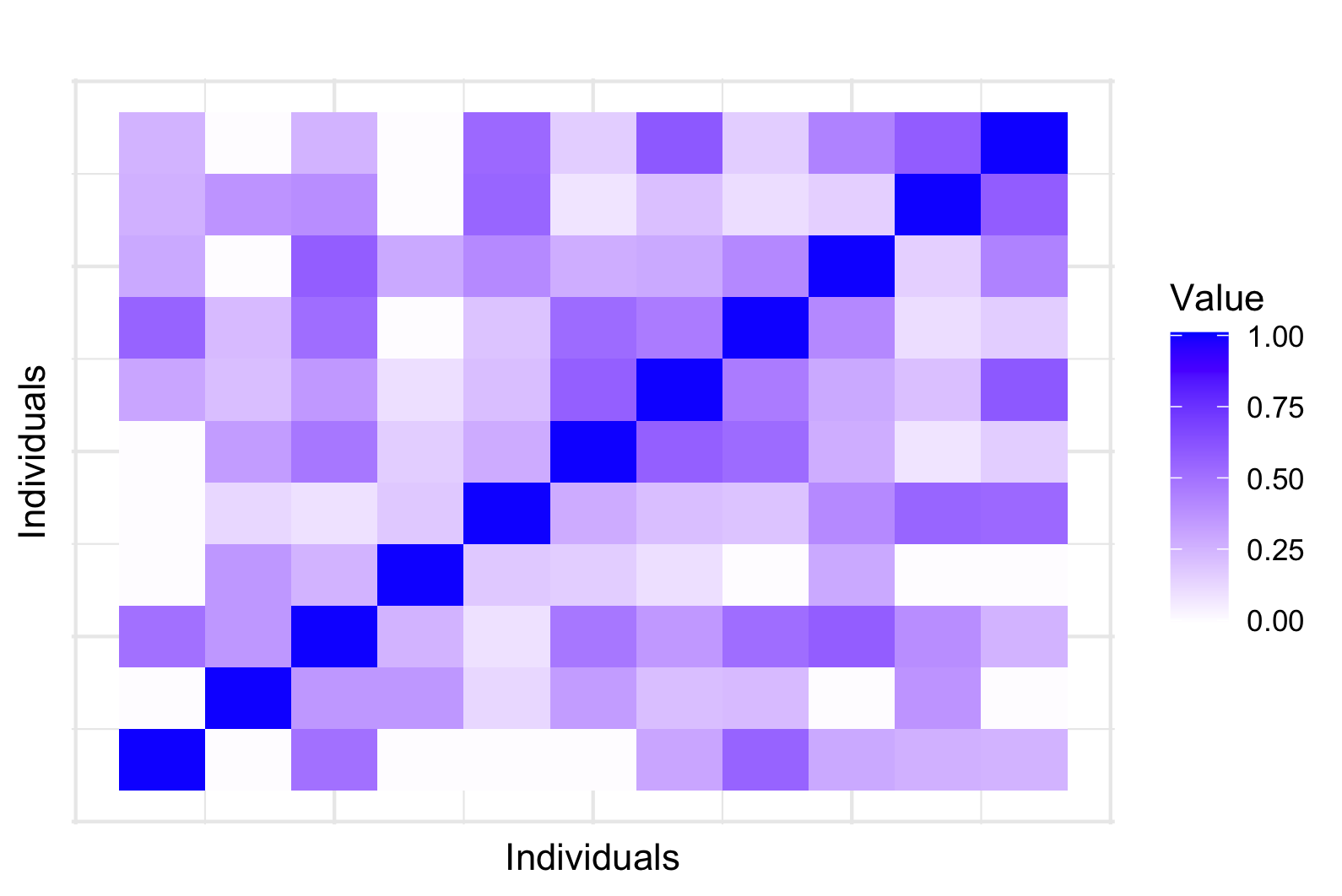}
    \caption{Degree kernel across individuals with comorbidity (both GAD and MDD). The matrix shows the pairwise kernel similarity value among individual networks.}
    \label{fig:kernel_individuals_comorbidity}
\end{figure}

Following our first pipeline (see Figure \ref{fig:pipeline}, left branch), we started the analysis of the extended dataset on GAD and MDD from \cite{fisher2017exploring} by applying the PCMCI+ algorithm to each individual’s time series, yielding 45 mixed causal graphs that incorporate both contemporaneous (lag-0) and lagged (lag-1) relationships, see the Supplementary Information, Section S2. Then, we quantified pairwise similarities between them using the degree distribution kernel to evaluate the extent of heterogeneity within each diagnostic category. Figures~\ref{fig:kernel_individuals_GAD}, ~\ref{fig:kernel_individuals_MDD} and ~\ref{fig:kernel_individuals_comorbidity} display the three similarity matrices (one per each diagnostic group), showing a substantial variability in the graph structure even among individuals sharing the same diagnosis. These findings support prior literature, highlighting within-diagnosis heterogeneity in symptom expression and network topology \parencite{hoekstra2023heterogeneity, fisher2018lack}.

\begin{figure}[H]
    \centering
    \includegraphics[width=\linewidth]{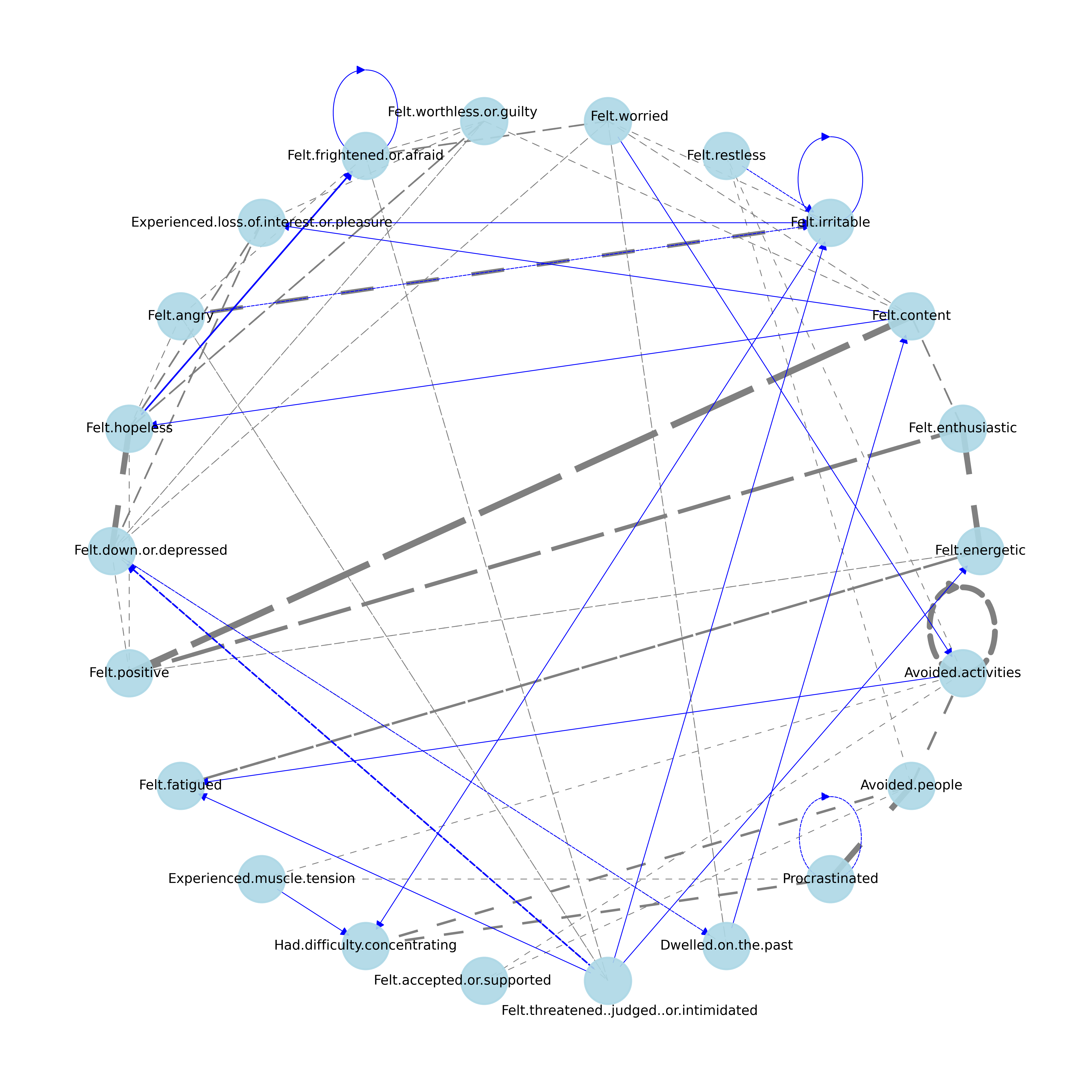}
    \caption{Fusion causal network for Generalized Anxiety Disorder (GAD). The graph represents the group-level symptom dynamics derived from individual causal networks. Grey and blue arrows indicate statistically significant undirected and directed causal relationships, respectively. Dashed and solid lines indicate lag-0 and lag-1 causal relationships, respectively.
    }
    \label{fig:fusion_networks_GAD}
\end{figure}

\begin{figure}[H]
    \centering
    \includegraphics[width=\linewidth]{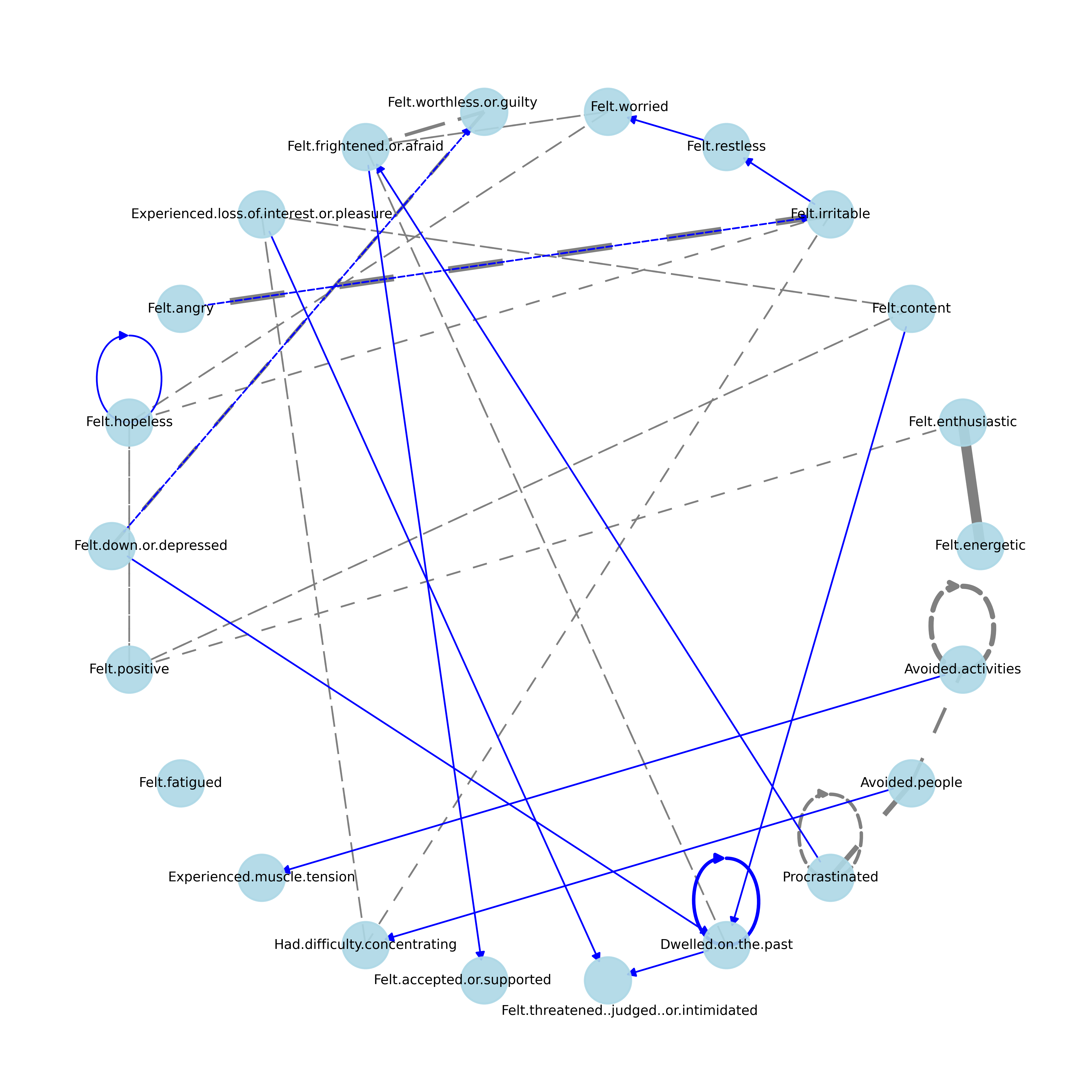}
    \caption{Fusion causal network for Major Depressive Disorder (MDD). The graph represents the group-level symptom dynamics derived from individual causal networks. Grey and blue arrows indicate statistically significant undirected and directed causal relationships, respectively. Dashed and solid lines indicate lag-0 and lag-1 causal relationships, respectively.}
    \label{fig:fusion_networks_MDD}
\end{figure}

\begin{figure}[H]
    \centering
    \includegraphics[width=\linewidth]{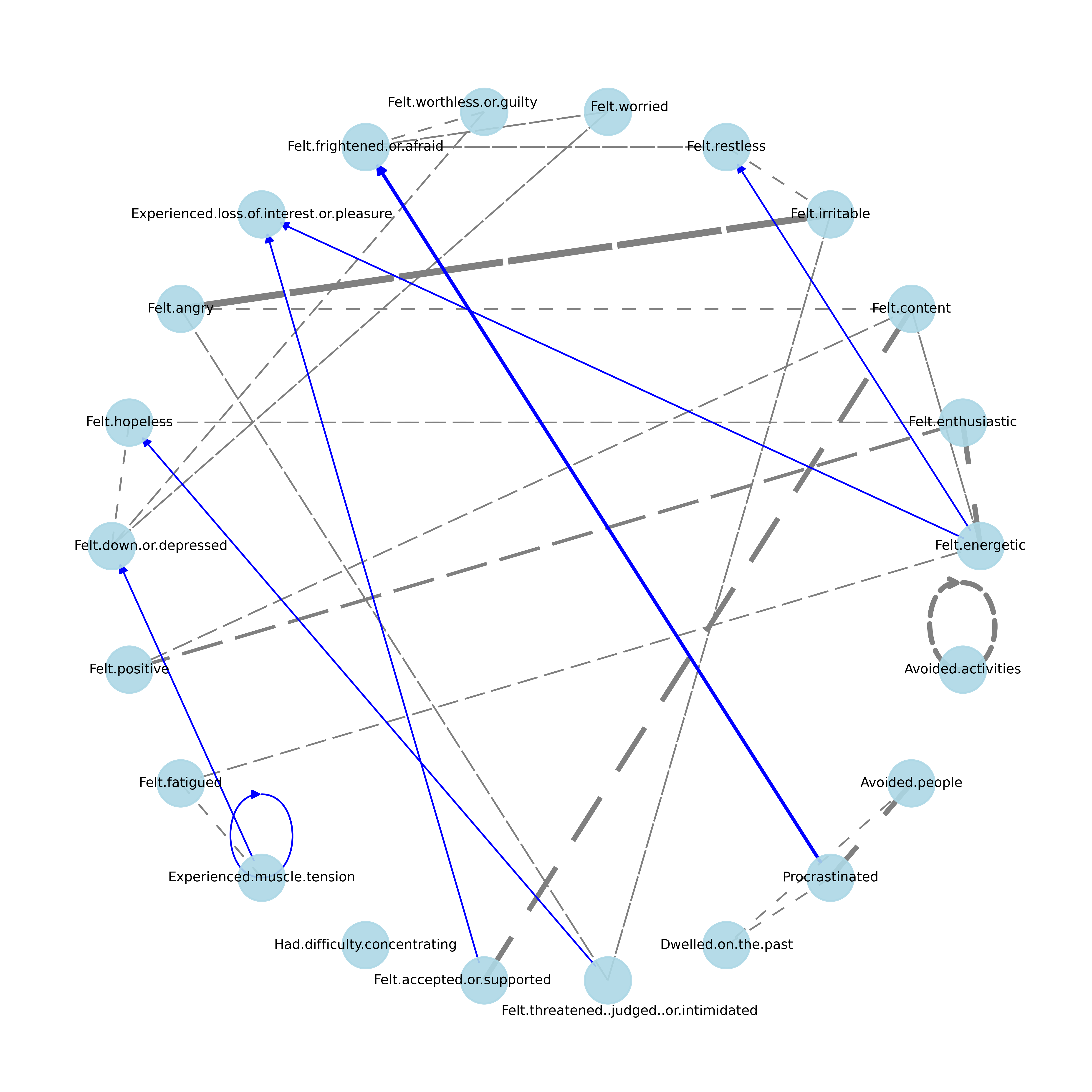}
    \caption{Fusion causal network for comorbidity (both GAD and MDD). The graph represents the group-level symptom dynamics derived from individual causal networks. Grey and blue arrows indicate statistically significant undirected and directed causal relationships, respectively. Dashed and solid lines indicate lag-0 and lag-1 causal relationships, respectively.}
    \label{fig:fusion_networks_comorbidity}
\end{figure}

To identify broader diagnostic patterns while preserving individual complexity, we considered the fusion causal networks, obtained by summing the individual adjacency matrices within each diagnostic group. This procedure yielded one group-level network for GAD, one for MDD, and one for comorbid GAD and MDD. Figures~\ref{fig:fusion_networks_GAD},~\ref{fig:fusion_networks_MDD} and ~\ref{fig:fusion_networks_comorbidity} display these full aggregated graphs.

\begin{figure}
    \centering
    \includegraphics[width=\linewidth]{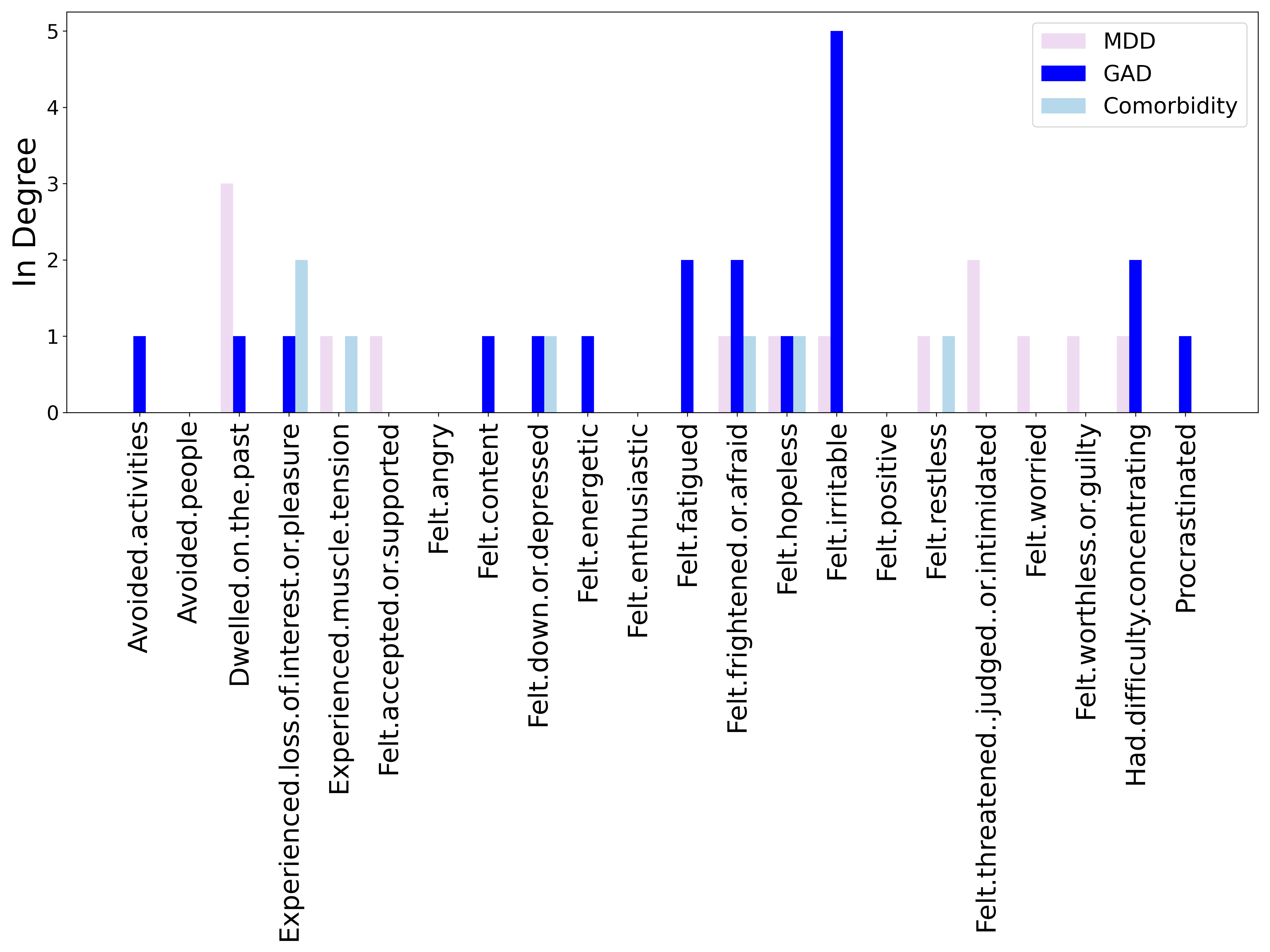}
    \caption{Comparison of in-degree values across symptoms for General Anxiety Disorder (GAD), Major Depressive Disorder (MDD) and comorbidity fusion networks. Each bar represents the number of incoming connections for a given symptom in the directed network.}
    \label{fig:in_degree}
\end{figure}

\begin{figure}
    \centering
    \includegraphics[width=\linewidth]{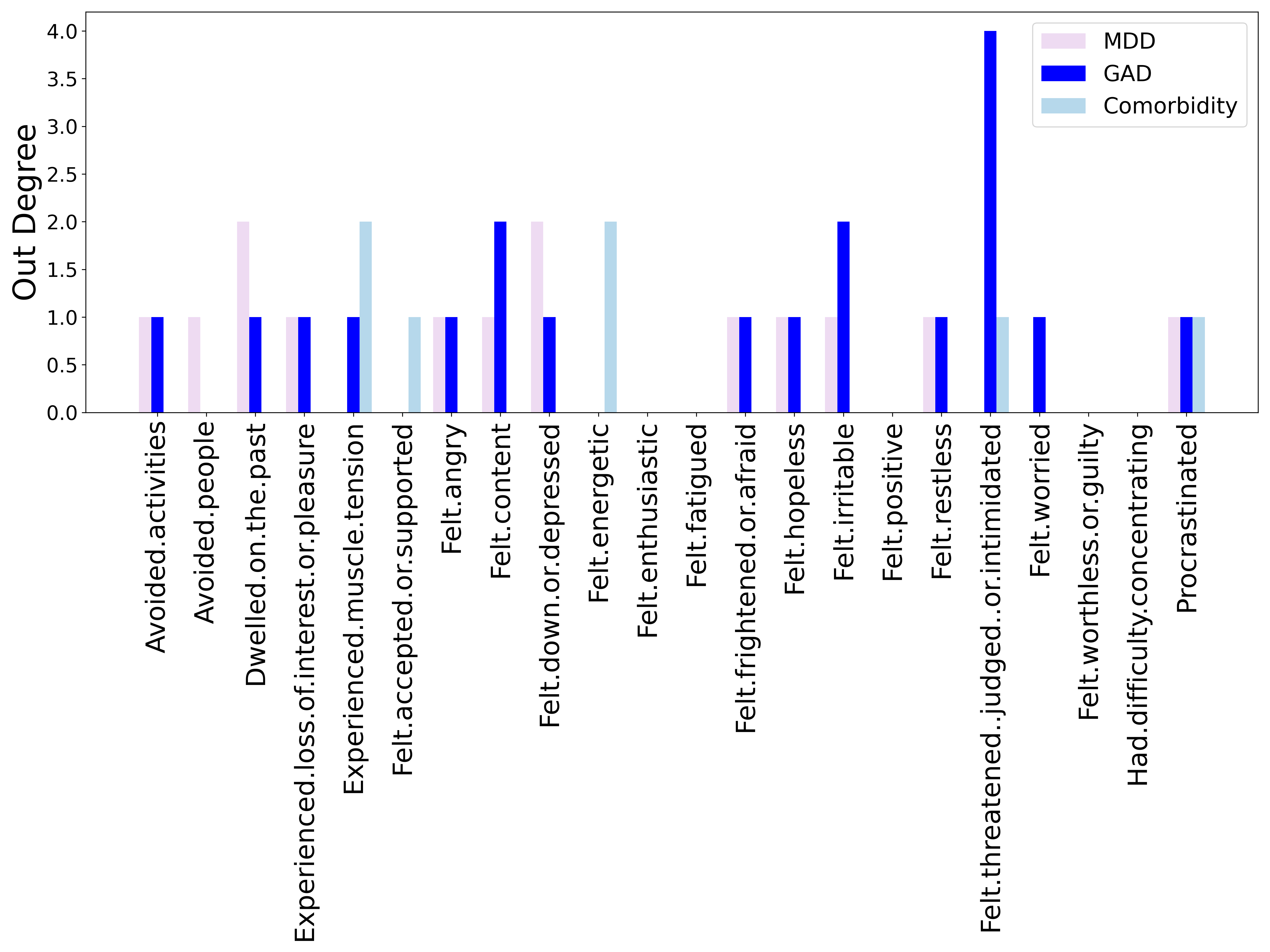}
    \caption{Comparison of out-degree values across symptoms for General Anxiety Disorder (GAD), Major Depressive Disorder (MDD) and comorbidity fusion networks. Each bar represents the number of outgoing connections for a given symptom in the directed network.}
    \label{fig:out_degree}
\end{figure}

\begin{figure}
    \centering
    \includegraphics[width=\linewidth]{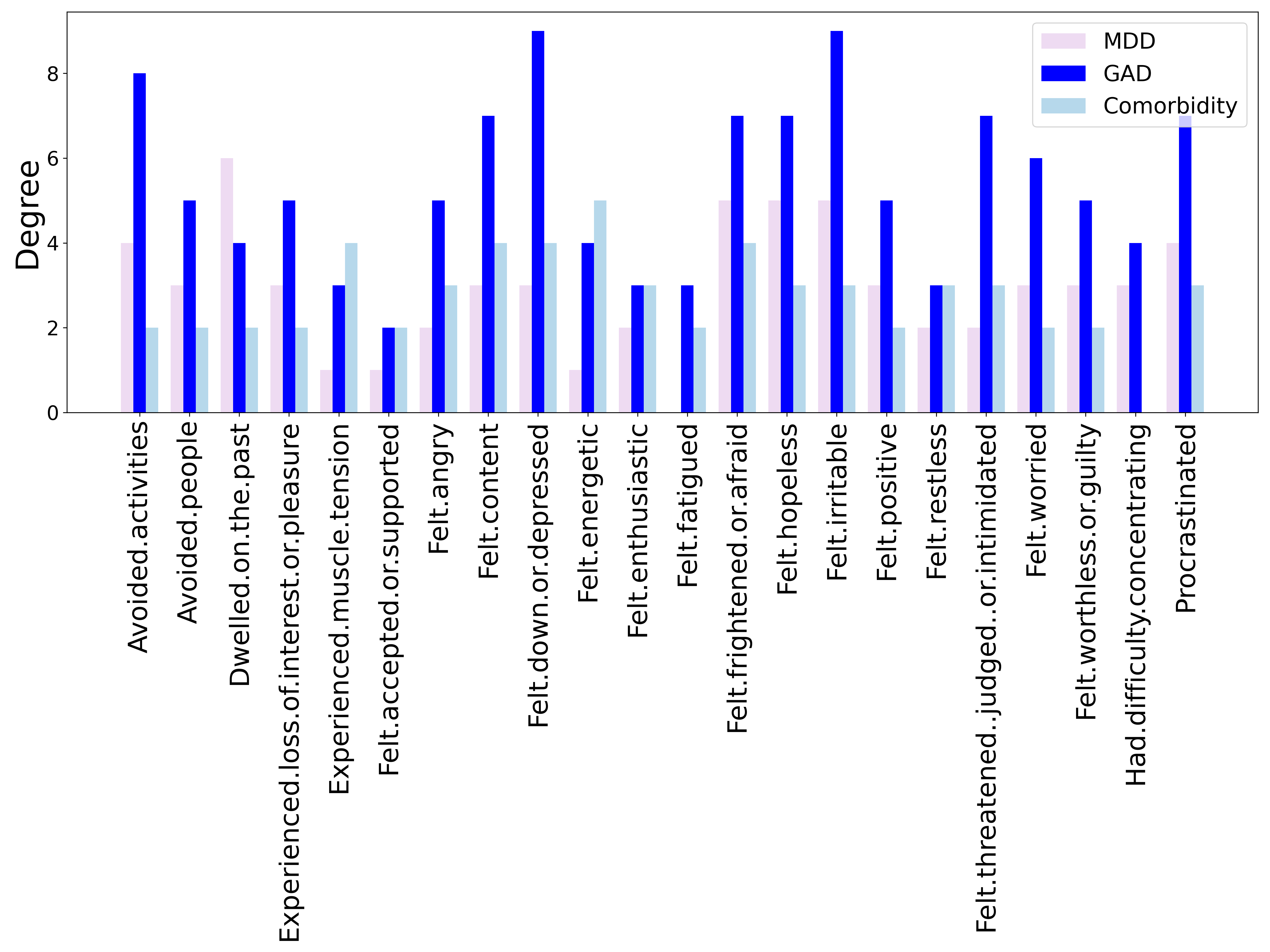}
    \caption{Comparison of degree values across symptoms for General Anxiety Disorder (GAD), Major Depressive Disorder (MDD) and comorbidity fusion networks. Each bar represents the number of connections for a given symptom in the mixed network.}
    \label{fig:degree}
\end{figure}

\begin{figure}
    \centering
    \includegraphics[width=\linewidth]{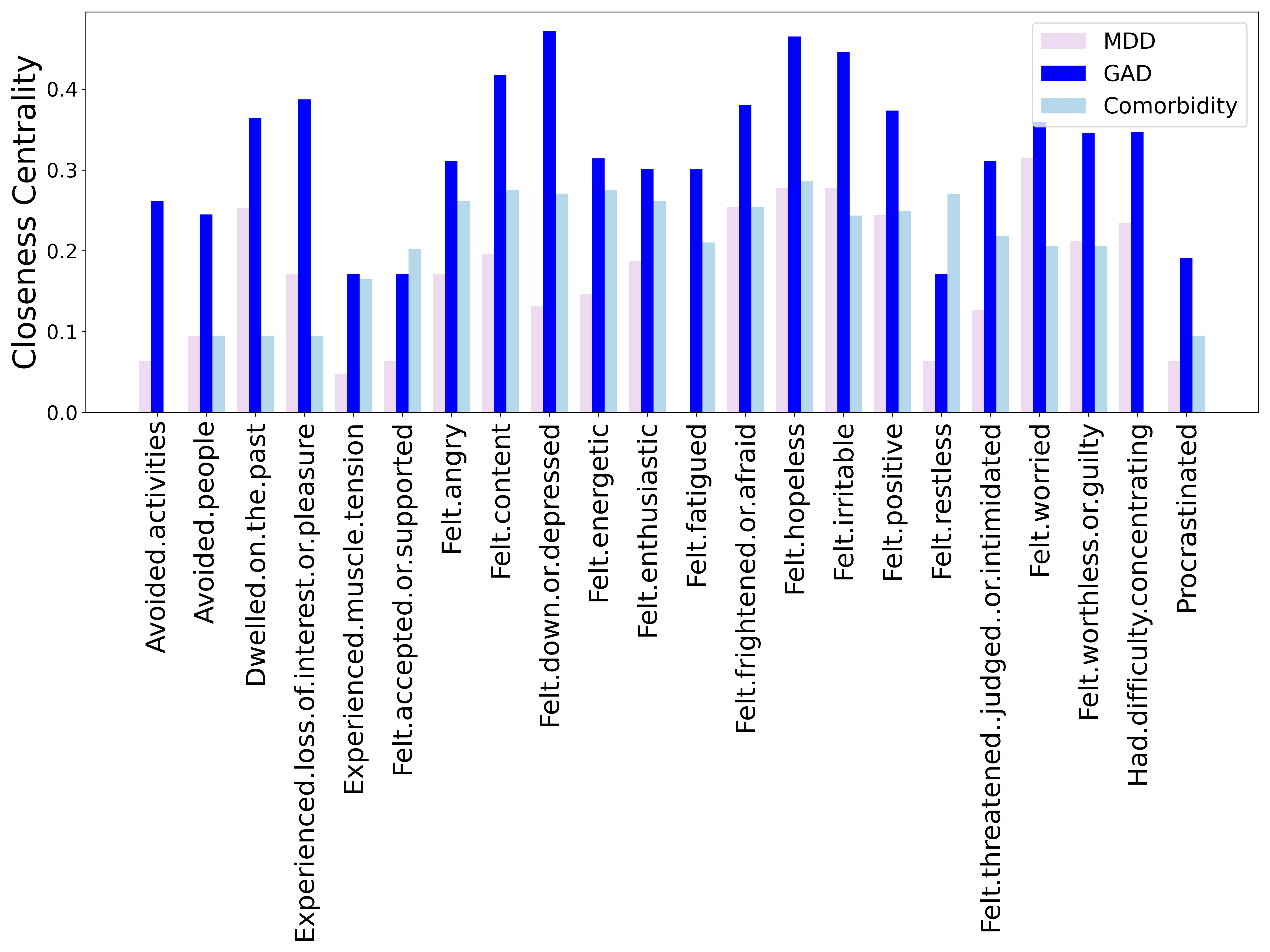}
    \caption{Comparison of closeness centrality values across symptoms for General Anxiety Disorder (GAD), Major Depressive Disorder (MDD) and comorbidity fusion networks.}
    \label{fig:closeness}
\end{figure}

\begin{figure}
    \centering
    \includegraphics[width=\linewidth]{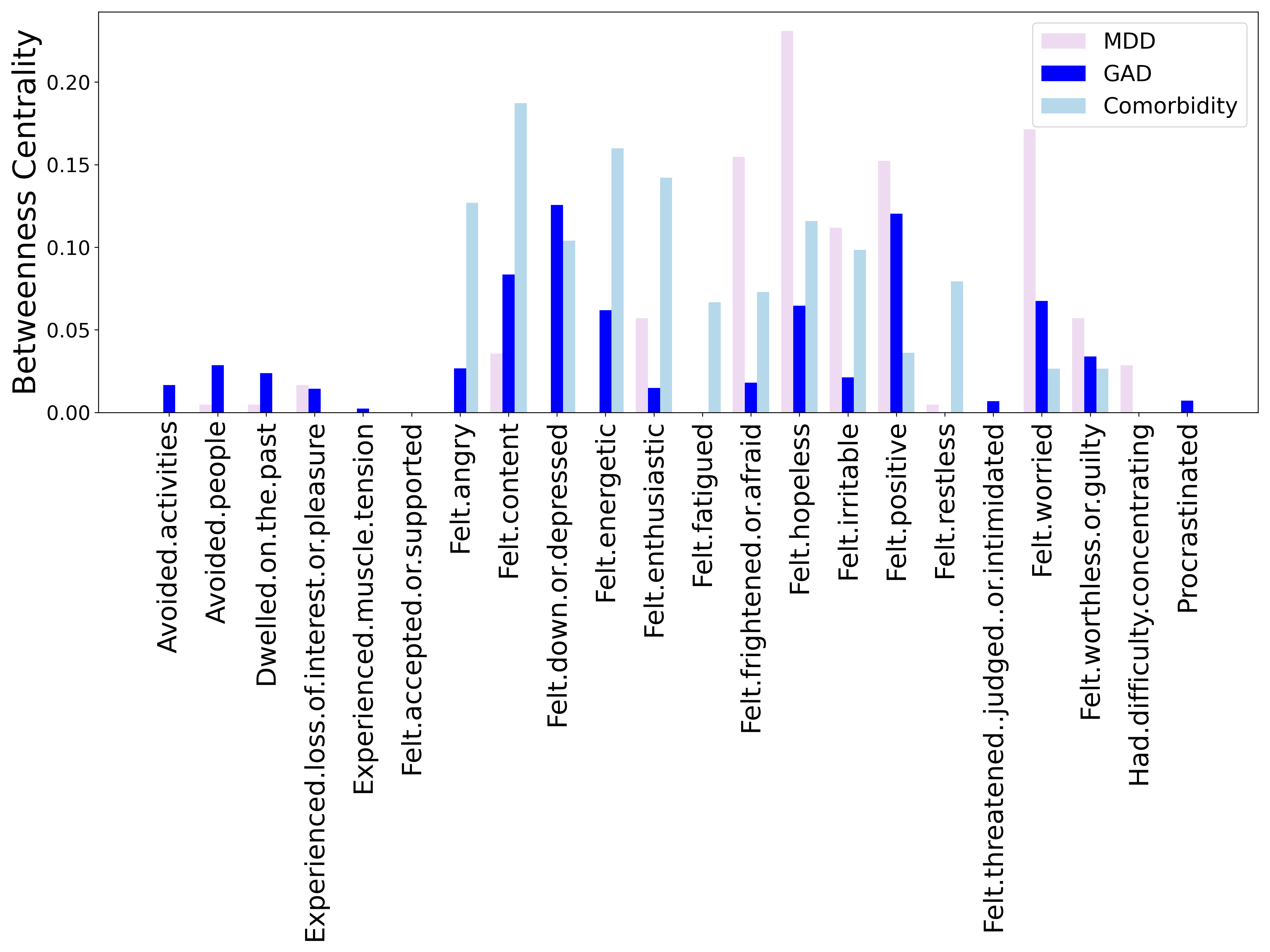}
    \caption{Comparison of betweenness centrality values across symptoms for 
    General Anxiety Disorder (GAD), Major Depressive Disorder (MDD) and comorbidity fusion networks.}
    \label{fig:betweenness}
\end{figure}

A visual inspection of the graphs---supported by quantitative network measures, see the Supplementary Information, Section S3 for details--- highlights significant differences in causal interactions between diagnostic groups, especially between GAD and MDD (see Figures \ref{fig:in_degree}---\ref{fig:betweenness}). 

Notably, MDD and comorbid networks show higher values of betweenness centrality, while GAD is characterized by higher degrees and closeness centrality. This pattern aligns with the literature that conceptualize GAD as characterized by physiological hyperarousal and pervasive cognitive worry, leading to highly interconnected and rapidly interacting symptoms \parencite{nitschke2001distinguishing}. In contrast, the more compartmentalized network structures observed in MDD and comorbid cases—with central symptoms acting as mediators—are consistent with the literature emphasizing reduced behavioral engagement, narrowed affective experience, and distinct symptom clusters \parencite{barlow1991nature, watson1988positive}.

We also observe several symptom-specific differences among diagnostic groups. For instance, in the GAD network, the symptom “dwelled on the past” functions as a driver, influencing “felt content”. In contrast, in MDD, it acts more as an endpoint, often reinforced by self-loops: indeed, it has the highest in-degree value for MDD (Figure \ref{fig:in_degree}). Similarly, “felt irritable” plays a central role in GAD, strongly influencing several other symptoms and having the highest in-degree in that network, while its role is more marginal in MDD.

The symptoms “felt hopeless” and “felt down or depressed” exhibit the highest closeness centrality in GAD, see Figure \ref{fig:closeness}, and are also connected by the second strongest undirected link. This is not true for MDD, where the two symptoms are even disconnected. “Felt threatened, judged, or intimidated” further exemplifies differential dynamics in the two disorders, whereby it is a cascade initiator in GAD, showing the highest out-degree (Figure \ref{fig:out_degree}), whereas in MDD, it emerge as a consequence of other symptoms and has the second-highest in-degree. More generally, we can identify a few instances, where initiating symptoms in MDD become terminal symptoms in GAD or viceversa.


The comorbid GAD and MDD network, while sharing elements with both GAD and MDD, also reveals unique patterns. For instance, “dwelled on the past” is linked to “procrastinated” and “avoided people,” suggesting a pattern of avoidance and disengagement not as clearly observed in the other groups. Additionally, “experienced muscle tension” appears to precede “felt down or depressed,” potentially indicating a stronger somatic contribution to mood in the comorbid profile. Interestingly, “experienced loss of interest or pleasure” serves as an arrival point in this network, in contrast to its trigger role in GAD and MDD (Figure \ref{fig:in_degree}).

Notably, in the comorbid GAD and MDD network, “felt content” exhibits the highest betweenness centrality across all groups, acting as a key bridge in the flow of symptom activation.
Overall, these patterns suggest that comorbidity does not simply reflect an additive overlap of GAD and MDD symptoms but rather reveals unique and potentially more complex causal interactions. 

\begin{figure}
    \centering
    \includegraphics[width=\linewidth]{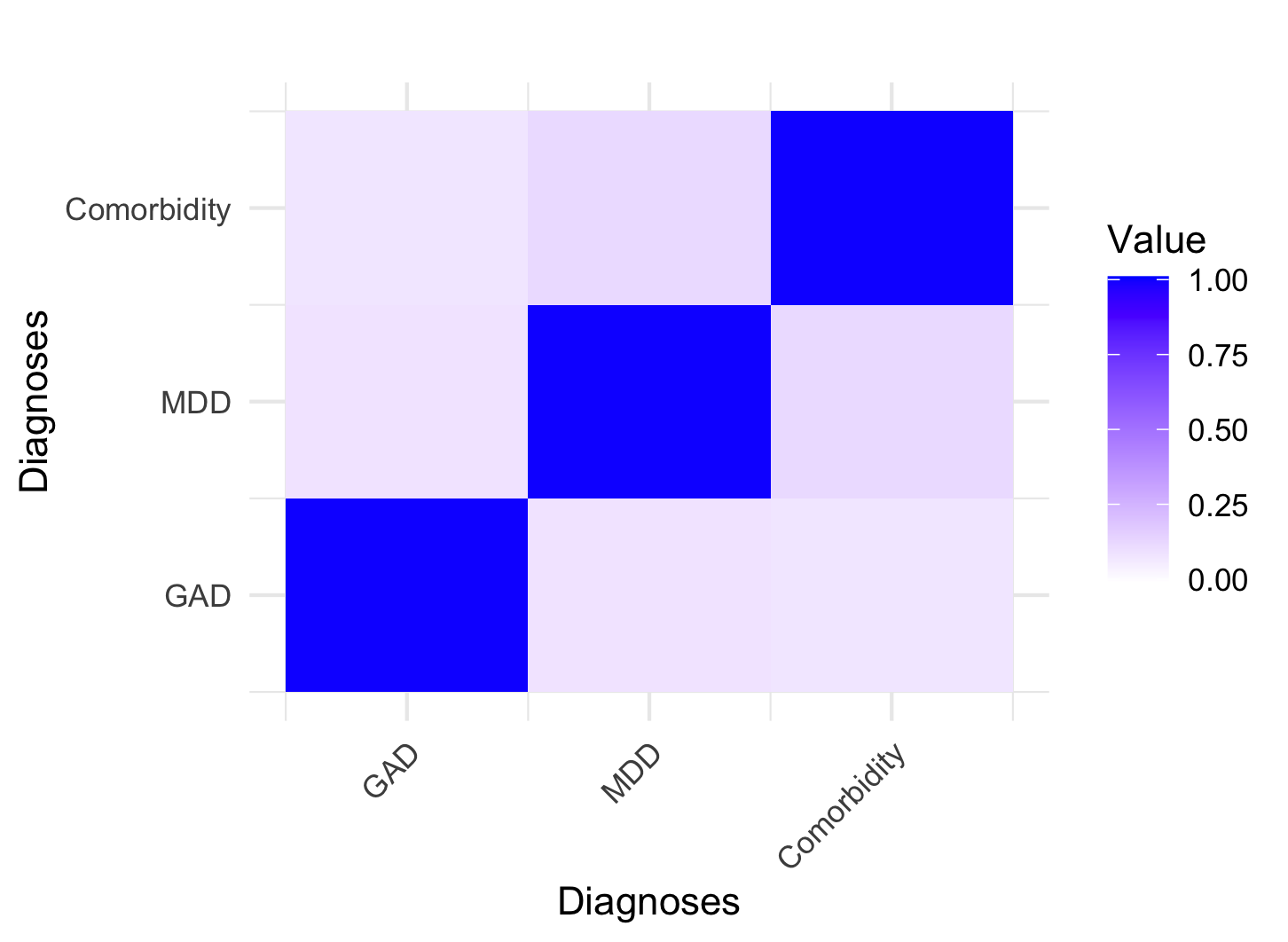}
    \caption{Weisfeiler-Lehman (WL) graph kernel similarity matrix comparing the fusion causal networks for Generalized Anxiety Disorder (GAD), Major Depressive Disorder (MDD), and comorbid cases (GAD and MDD). The matrix shows the pairwise kernel similarity value among diagnostic groups.}
    \label{fig:kernel_groups}
\end{figure}

Our findings highlight how symptom interactions—not just symptom presence—may differ between disorders. This qualitative observation is further backed by  the Weisfeiler-Lehman (WL) graph kernel used to assess the similarity between the aggregated causal networks for GAD, MDD, and comorbid cases. Indeed, the resulting similarity matrix revealed a strong and clear separation across groups, see Figure~\ref{fig:kernel_groups}. This suggests that diagnostically meaningful structure can be recovered when individual networks are aggregated, despite substantial intra-group variability.

\subsection{Complexity Measures are Key towards an Accurate Diagnostic Classification.}

The application of the PCMCI+ algorithm revealed structural differences in the causal networks describing the three diagnostic groups. Our second pipeline (see Figure \ref{fig:pipeline}, right branch) now aims at predicting diagnosis based on symptom dynamics, which is a critical goal for clinical purposes. Following our pipeline, we enriched the dataset by computing a set of complexity-based features extracted from each individual’s symptom time series (see Table~\ref{tab:complexity_metrics} and the Supplementary Information, Section S4 for their definitions), which helped extract the salient dynamical aspects of symptom expression.

\begin{figure}
    \centering
    \includegraphics[width=\linewidth]{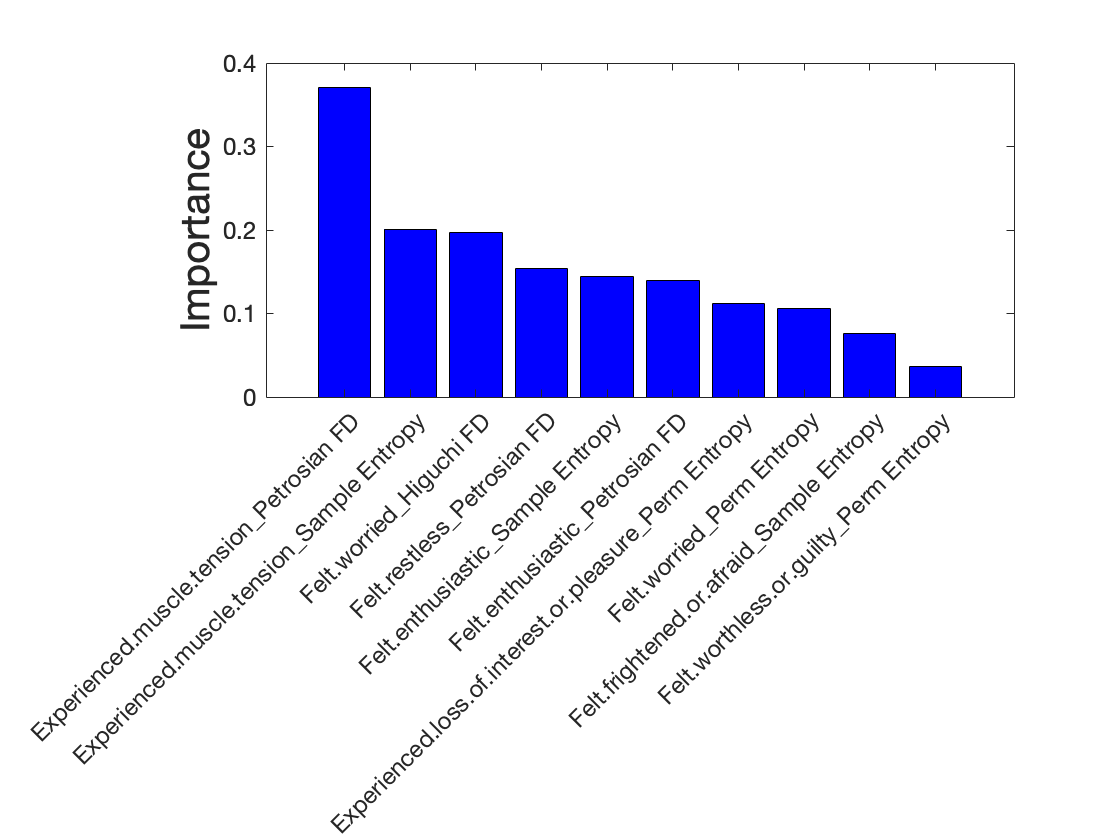}
    \caption{Most important features identified as crucial in distinguishing between General Anxiety Disorder (GAD) and Major Depressive Disorder (MDD) based on feature importance scores from the Bagged Tree classifier.}
    \label{fig:feature_importance}
\end{figure}

\begin{figure}
    \centering
    \includegraphics[width=\linewidth]{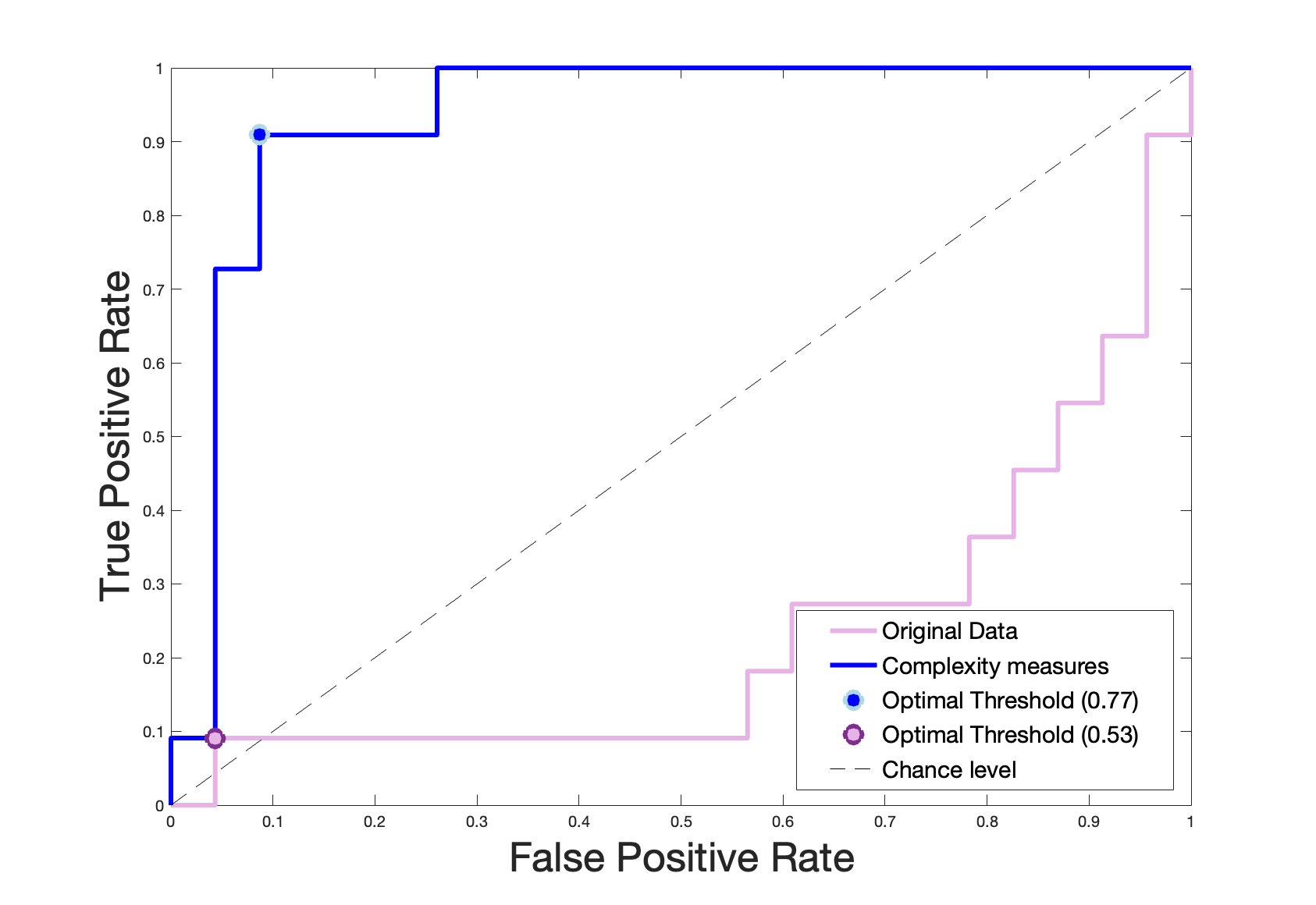}
    \caption{Receiver Operating Characteristic (ROC) curve for the classification between General Anxiety Disorder (GAD) and Major Depressive Disorder (MDD). The blue (violet) curve includes the selected complexity-based features (the original data). Each curve illustrates the model's performance across classification thresholds, with an Area Under the Curve (AUC) of 0.92 (0.23). The blue (violet) circle marks the optimal classification threshold of 0.77 (0.53), above which cases are classified as MDD. These thresholds were selected to maximize both sensitivity and specificity. The diagonal dashed line represents chance-level performance.}
    \label{fig:roc_curve}
\end{figure}

Next, starting from this enriched dataset, we performed feature selection using Boruta-type out-of-bag (OOB) variable importance scores to improve interpretability and filter out noise \parencite{kursa2010feature}. The trained classifier obtained by involving only the 10 top-ranked features shown in Figure~\ref{fig:feature_importance}, selected through the iterative procedure described in Section~\nameref{sec:methods} yields excellent performance: AUC $= 0.92$, sensitivity $= 0.91$, specificity $= 0.91$, accuracy $= 0.91$. We emphasize that performance is substantially higher compared to that of the Bagged Tree ensemble model trained on the raw symptom time series (AUC $= 0.23$, sensitivity $= 0.09$, specificity $= 0.91$, accuracy $= 0.65$), see Figure \ref{fig:roc_curve} for a detailed comparison through ROC curves. The substantial performance improvement is likely due to the nature of the selected  features, which include entropy and fractal-based descriptors of key symptoms, thereby reflecting the multidimensional and dynamic nature of symptoms across diagnoses. 

Finally, we assessed the generalizability of the classificator by applying the  model to the 11 participants with comorbid GAD and MDD, who were excluded from training. Interestingly, the classifier produced predictions that aligned with each participant’s dominant clinical profile. In particular, when HAM-A scores exceeded HAM-D scores, the predicted diagnosis was typically GAD, and vice versa. A summary of these predictions is presented in Table~\ref{tab:comorbidity_predictions}.

\begin{table}[ht]
\centering
\begin{threeparttable}
\caption{Model predictions for comorbid participants, not included in training. The model assigns diagnosis based on a probability threshold of 0.77 for classifying MDD, as determined by the leave-one-out cross-validation.}
\label{tab:comorbidity_predictions}
\begin{tabular}{lcccc}
\toprule
\textbf{ID} & \textbf{Predicted Class} & \textbf{Pr(MDD)} & \textbf{HAM-D} & \textbf{HAM-A} \\
\midrule
P001  & GAD & 0.663 & 23 & 27 \\
P003  & GAD & 0.560 & 16 & 15 \\
P006  & GAD & 0.473 & 13 & 13 \\
P007  & GAD & 0.043 & 11 & 17 \\
P008  & GAD & 0.153 & 19 & 15 \\
P010  & MDD & 0.866 & 22 & 22 \\
P013  & GAD & 0.273 & 14 & 19 \\
P048  & GAD & 0.570 & 14 & 17 \\
P115  & GAD & 0.713 & 18 & 19 \\
P117  & GAD & 0.463 & 12 & 18 \\
P203  & GAD & 0.253 & 18 & 20 \\
\bottomrule
\end{tabular}
\begin{tablenotes}
\small
\item Note. HAM-D = Hamilton Rating Scale for Depression score; HAM-A = Hamilton Rating Scale for Anxiety score; MDD = major depressive disorder; GAD = generalized anxiety disorder.
\end{tablenotes}
\end{threeparttable}
\end{table}

\section{Discussion}\label{sec:discussion}
This study aimed to enhance our understanding of symptom dynamics in clinical psychology, integrating advanced causal inference techniques, network-based methods, and machine learning tools. Our novel, integrated approach has been demonstrated on a case study where we attempt at deciphering the specificities of Generalized Anxiety Disorder (GAD), Major Depressive Disorder (MDD), and their comorbidity. Addressing a critical gap in previous literature—specifically, the tension between individual-level complexity and group-level diagnostic generalization—we implemented a structured analytical pipeline with three core methodological innovations.

Firstly, applying the PCMCI+ causal discovery algorithm to our case study on GAD and MDD participants revealed meaningful causal structures within individual symptom trajectories. Unlike traditional models (e.g., VAR and Transfer Entropy), PCMCI+ is capable of effectively identifying both linear and nonlinear causal relationships from short time series.

Secondly, by employing fusion networks combined with graph kernel methods to the PCMCI+ results, we successfully demonstrated that meaningful group-level patterns could emerge despite significant individual heterogeneity. The Weisfeiler-Lehman kernel highlighted clear structural differences between diagnostic groups, validating the potential of aggregating individual causal graphs into coherent, group-level representations. This methodological step provides a robust view of diagnostic differences at a structural level.

Importantly, our findings align well with previous theoretical and clinical conceptualizations \parencite{eysenck2006anxiety, nitschke2001distinguishing}, demonstrating that GAD and MDD indeed differ at the causal level, particularly concerning central symptoms such as rumination, avoidance, and emotional reactivity.

Thirdly, incorporating complexity measures within a machine learning framework—specifically, a Bagged Tree classifier—substantially enhanced diagnostic classification accuracy compared to models relying only on raw symptom data. Notably, entropy, fractal dimensions, and recurrence-based metrics proved crucial in distinguishing between GAD and MDD, and to inform on the prevalent diagnosis in comorbidity cases. The final classification model, trained on selected complexity-based features, achieved an impressive accuracy ($\geq 0.9$), underscoring the clinical relevance of dynamic complexity in symptom expression. To the best of our knowledge, the use of complexity measures represents a novel extension in clinical psychology aimed at exploring whether complex temporal features can help differentiate diagnostic categories based on symptom dynamics.

The findings of this manuscript pave the way for fundamental advances in our understanding of mental disorders, whereby they represent a solid step towards discriminating disorders by means of their symptom dynamics. Indeed, the first pipeline, grounded on the PCMCI+ algorithm, allows to discriminate the main group-level differences in symptom dynamics across mental disorders, filtering out the unavoidable individual heterogeneities. The second pipeline then allows the identification of the complexity measures that best capture the differential symptom dynamics across psychopathologies.

In addition, our results seem promising also towards personalized diagnostics and therapeutic interventions. Indeed, the outcome of the two pipelines may prove to be a valid support for practitioners. The classification coming from the second pipeline can represent a first, automatic screening supporting the diagnosis. In a prudent approach, the automatic diagnosis can represent an alert flag for the practitioner. Once a diagnosis has been finalized, the individual causal graphs from the first pipeline can be then used to support personalized therapy, whereby it allows to identify central and source nodes, which may prove critical for the persistence of the disorder.

Despite these advances, our study is not free of limitations, which should be addressed in future work. First, our pipelines should be tested on other datasets, to further stress the ability of our approach to deal with short time series and relatively small sample sizes. Second, towards the application of the pipelines to support practitioners in diagnosis, datasets including control groups should also be tested. Moreover, from a methodological standpoint, one should consider that the PCMCI+ algorithm—despite its strengths—relies on several key assumptions. In particular, it assumes causal sufficiency, meaning that all relevant variables influencing the system have been measured. This implies that if hidden confounders are present, the inferred causal relations may be biased or spurious. Moreover, while PCMCI+ can capture both linear and nonlinear dependencies, it may still be sensitive to violations of temporal stationarity and to the quality of the time series data, and therefore suitable data preprocessing might be required \parencite{ryan2025non}.
Finally, an underlying assumption in graph representations is that symptom interactions are pairwise. However, recent work has pointed out that higher-order multibody interactions between symptoms may play a role in explaining their dynamics. Alternative metrics that also account for higher-order interaction, such as the O-information, could be considered \parencite{marinazzo2024information}.

\section*{Acknowledgments}

We would like to thank Professor Aaron J. Fisher for making the dataset publicly available and for his pioneering work on idiographic modeling in clinical psychology, which greatly inspired our approach.

\section*{Declarations}
\paragraph{Funding}
This work was supported by the European Union, the Next Generation EU project ECS00000017 ‘Ecosistema dell’Innovazione’ Tuscany Health Ecosystem (THE, PNRR, Spoke 4: Spoke 9: Robotics and Automation for Health).

\paragraph{Conflicts of interest} The authors have no financial or proprietary interests in any material discussed in this article.

\paragraph{Ethics approval} Not applicable.

\paragraph{Consent to participate} Not applicable.

\paragraph{Consent for publication} Not applicable.

\paragraph{Availability of data and materials}
 The data of this work is available at \url{https://github.com/elevitanz/complex_psychology}

\paragraph{Code availability}
 The code of this work is available at
\url{https://github.com/elevitanz/complex_psychology}

\printbibliography
\end{document}